\documentclass[aps,prl,twocolumn,amsmath,amssymb,showpacs, superscriptaddress,notitlepage,longbibliography]{revtex4-1}

\usepackage[colorlinks=true,linkcolor=blue,anchorcolor=red,citecolor=blue, urlcolor=blue]{hyperref}
\usepackage{bm}
\usepackage{graphicx}
\usepackage{color}
\usepackage{cases}

\begin{document}

\title{Rules for Phase Shifts of Quantum Oscillations in Topological Nodal-line Semimetals}

\author{Cequn Li}

\affiliation{Shenzhen Institute for Quantum Science and Engineering and Department of Physics, Southern University of Science and Technology, Shenzhen 518055, China}
\affiliation{Shenzhen Key Laboratory of Quantum Science and Engineering, Shenzhen 518055, China}
\affiliation{Department of Physics, The Pennsylvania State University, University Park, Pennsylvania 16802, USA}

\author{C. M. Wang}

\affiliation{Shenzhen Institute for Quantum Science and Engineering and Department of Physics, Southern University of Science and Technology, Shenzhen 518055, China}

\affiliation{Shenzhen Key Laboratory of Quantum Science and Engineering, Shenzhen 518055, China}

\affiliation{School of Physics and Electrical Engineering, Anyang Normal University, Anyang 455000, China}

\author{Bo Wan}
\affiliation{National Laboratory of Solid State Microstructures, Collaborative Innovation Center of Advanced Microstructures, School of Physics, Nanjing University, Nanjing 210093, P. R. China}
\affiliation{Shenzhen Institute for Quantum Science and Engineering and Department of Physics, Southern University of Science and Technology, Shenzhen 518055, China}

\author{Xiangang Wan}
\affiliation{National Laboratory of Solid State Microstructures, Collaborative Innovation Center of Advanced Microstructures, School of Physics, Nanjing University, Nanjing 210093, P. R. China}

\author{Hai-Zhou Lu}
\email{Corresponding author. luhz@sustc.edu.cn}

\affiliation{Shenzhen Institute for Quantum Science and Engineering and Department of Physics, Southern University of Science and Technology, Shenzhen 518055, China}

\affiliation{Shenzhen Key Laboratory of Quantum Science and Engineering, Shenzhen 518055, China}

\author{X. C. Xie}
\affiliation{International Center for Quantum Materials, School of Physics, Peking University, Beijing 100871, China}
\affiliation{Collaborative Innovation Center of Quantum Matter, Beijing 100871, China}

\date{\today }
\begin{abstract}
Nodal-line semimetals are topological semimetals in which band touchings form nodal lines or rings. Around a loop that encloses a nodal line, an electron can accumulate a nontrivial $\pi$ Berry phase, so the phase shift in the Shubnikov-de Haas (SdH) oscillation may give a transport signature for the nodal-line semimetals. However, different experiments have reported contradictory phase shifts, in particular, in the WHM nodal-line semimetals (W=Zr/Hf, H=Si/Ge, M=S/Se/Te). For a generic model of nodal-line semimetals, we present a systematic calculation for the SdH oscillation of resistivity under a magnetic field normal to the nodal-line plane. From the analytical result of the resistivity, we extract general rules to determine the phase shifts for arbitrary cases and apply them to ZrSiS and Cu$_3$PdN systems. Depending on the magnetic field directions, carrier types, and cross sections of the Fermi surface, the phase shift shows rich results, quite different from those for normal electrons and Weyl fermions. Our results may help exploring transport signatures of topological nodal-line semimetals and can be generalized to other topological phases of matter.
\end{abstract}


\maketitle

\begin{figure}[tbph]
\centering \includegraphics[width=0.45\textwidth]{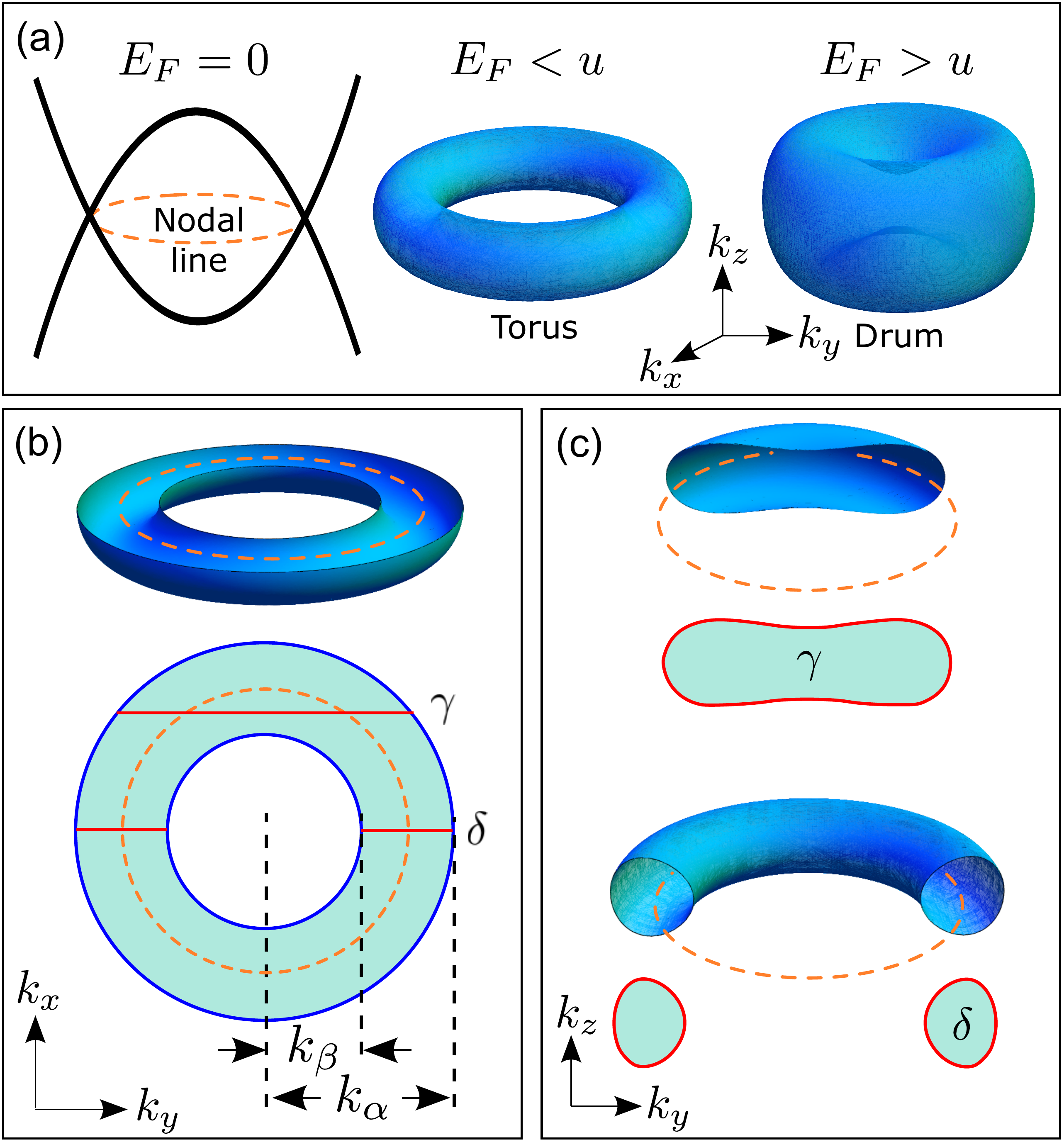}
\caption{(a) The nodal line, torus, and drum Fermi surfaces for a generic model of nodal-line semimetal in Eq. \eqref{eq:model}. The dashed ring stands for the nodal ring. $E_F$ is the Fermi energy. $u$ is a model parameter. (b) The maximum ($\alpha$) and minimum ($\beta$) cross sections in the nodal-line plane of the torus Fermi surface. (c) The maximum ($\gamma$) and minimum ($\delta$) cross sections out of the nodal-line plane of the torus Fermi surface.}
\label{fig:torus}
\end{figure}

{\color{red}\emph{Introduction.}}-- Nodal-line semimetals \cite{Burkov11prb,Chiu14prb,Fang16cpb,Yang17prb} are topological semimetals in which conduction and valence bands touch at open lines or closed rings  [Fig. \ref{fig:torus}(a)] in momentum space \cite{Chen15nl,Bzduvsek16nature}. Compared to topological insulators and Weyl semimetals, the nodal-line semimetals have unique topologically-constrained drumhead surface states. With broken symmetries, the nodal-line semimetals can evolve into various topological phases of matter, such as Dirac semimetals and topological insulators. Therefore, the nodal-line semimetals are of great fundamental interests. The nodal lines have been proposed in HgCr$_2$Se$_4$ \cite{Xu11prl}, graphene networks \cite{weng2015topological}, Cu$_3$(Pd/Zn)N \cite{yu2015topological,kim2015dirac}, SrIrO$_3$ \cite{fang2015topological,Chen2015nc}, TlTaSe$_2$ \cite{bian2016drumhead}, Ca$_3$P$_2$ \cite{xie2015new,chan20163}, CaTe \cite{Du17npjqm}, compressed black phosphorus \cite{zhao2016topological}, CaAg(P/As) \cite{yamakage2015line}, CaP$_3$ family \cite{xu2017topological}, PdS monolayer \cite{jin2017prediction}, Zintl compounds \cite{Zhu16prb}, BaMX$_3$ (M=V, Nb and Ta, X=S, Se) \cite{Liang16prb}, rare earth monopnictides \cite{zeng2015topological}, alkaline-earth compounds \cite{Hirayama17nc,Huang16prb,Li16prl}, other carbon based materials \cite{Chen15nl,Wang16prlbody}, and metallic rutile oxides XO$_2$, (X=Ir, Os, Rd) \cite{SunY17prb}; and experimentally verified in PbTaSe$_2$ \cite{bian2016topological}, ZrSiS \cite{schoop2016dirac,neupane2016observation,ChenC17prb} and PtSn$_4$ \cite{wu2016dirac} by angle-resolved photoemission spectroscopy.

In a nodal-line semimetal, an electron can pick up a nontrivial $\pi$ Berry phase around a loop that interlocks with the nodal line, while around a loop parallel to the nodal ring plane the Berry phase is trivial \cite{fang2015topological}. As a signature of the nodal-line semimetal, the nontrivial Berry phase may be probed by the quantum oscillations \cite{murakawa2013detection,WangCM16prl,He14prl,Novak15prbr,Zhao15prx,Du16scpma,YangXJ15arXiv-NbAs,Xiong15sci,Cao15nc,ZhangCL17np,Narayanan15prl,Park11prl,Xiang15prb,Tafti16np,Luo15prb}, that is,
the oscillation of the resistivity $\rho$ as a strong magnetic field quantizes the energy spectrum into Landau levels. The oscillation can be characterized by
$\rho \sim \cos[2\pi (F/B +\phi )]$,
where $B$ is the magnetic field strength, and the oscillation frequency $F$ and phase shift $\phi $ can provide the energy band and topological information of metals. In experiments, the phase shift of the quantum oscillation can be extracted from the Landau fan diagram, which presents the $n$th resistivity peak in perpendicular magnetic fields $B$ as a function of $1/B$. The slope of the linear fitting to the $n$-$1/B$ relation is the oscillation frequency and the intercept on the $n$-axis indicates the phase shift. Recently, quantum oscillations have been experimentally observed in ZrSiS \cite{singha2017large,ali2016butterfly,wang2016evidence,lv2016extremely,hu2016evidence1,PanH17arXiv}, ZrSi(Se/Te) \cite{hu2016evidence2}, ZrGe(S/Se/Te) \cite{hu2017quantum}, and HfSiS \cite{kumar2017unusual}. 
Singha \emph{et al.} \cite{singha2017large} and Ali \emph{et al}. \cite{ali2016butterfly} found that the phase shift of a higher frequency component is close to $\pm1/8$ or 0, implying a nontrivial Berry phase in 3D or 2D.
However, Wang {\it et. al.} \cite{wang2016evidence} claimed a trivial Berry phase for this higher frequency component in the same material ZrSiS.
To clarify the confusion, a theoretical study on the phase shift in nodal-line semimetals is highly desirable, and will be helpful for future exploration not only in nodal-line semimetals but also in other emergent topological materials.

\begin{figure}
\centering
\includegraphics[width=0.45\textwidth]{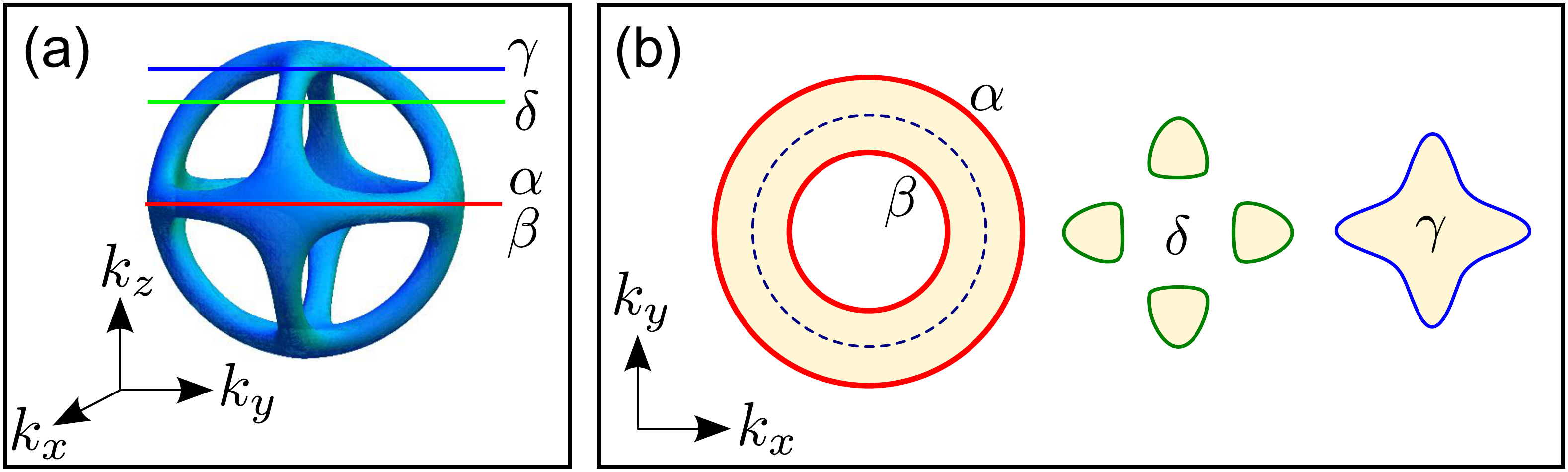}
\caption{(a) Fermi surface for the model of Cu$_3$PdN \cite{yu2015topological,kim2015dirac}. (b) The $\alpha$, $\beta$, $\gamma$, and $\delta$ extremal cross sections in (a).}\label{fig:CuPdN}
\end{figure}

\begin{figure}
\centering
\includegraphics[width=0.45\textwidth]{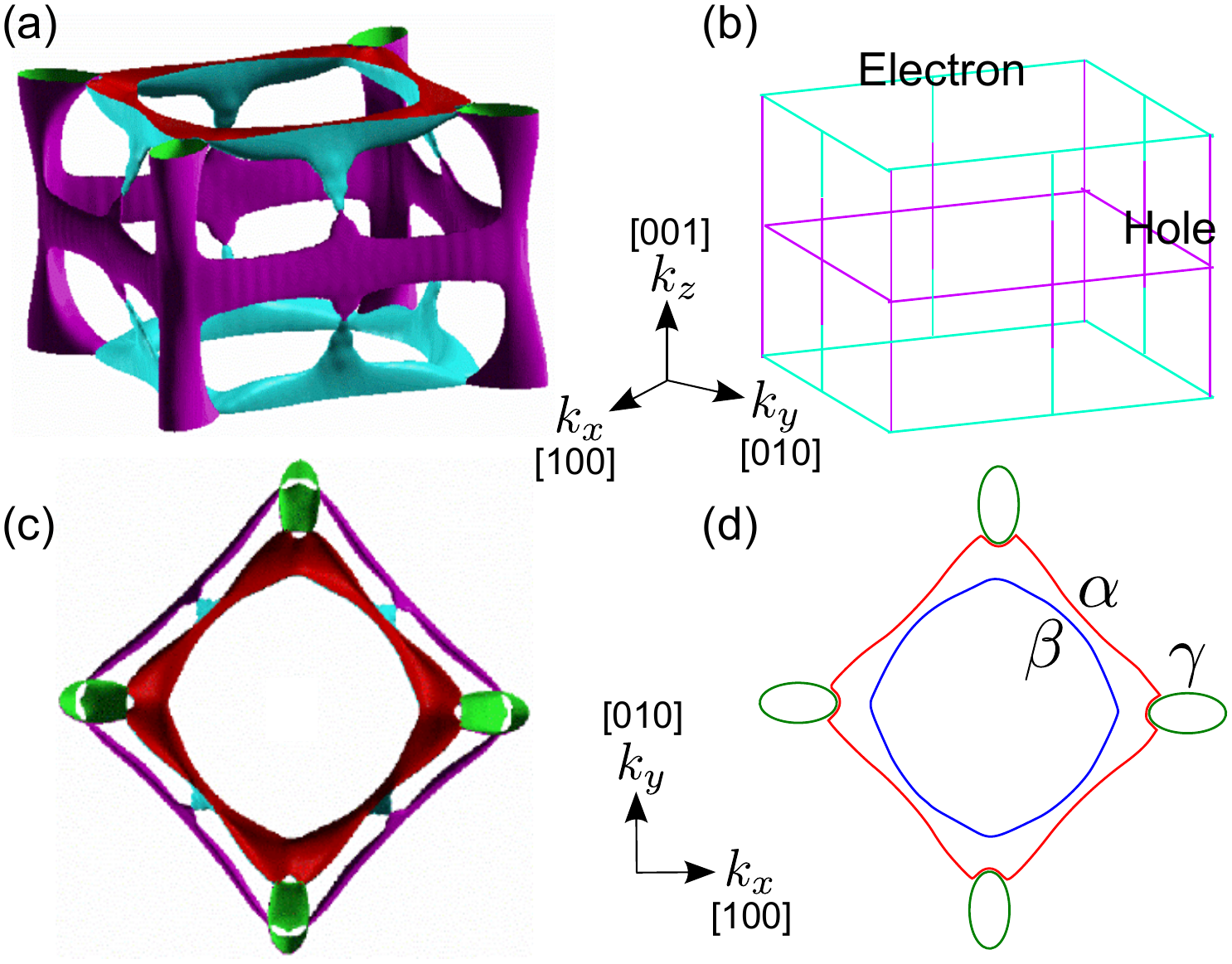}
\caption{(a) The Fermi surface of ZrSiS adapted from \cite{PanH17arXiv}. (b) The different colors distinguish the nodal-lines of electron and hole carriers, respectively. (c) Top view of (a). (d) The extremal cross sections in a [001]-direction magnetic field. }\label{fig:ZrSiS}
\end{figure}

In this Letter, we present a systematic calculation for the quantum oscillation of resistivity in topological nodal-line semimetals. We start with a generic model which describes a closed nodal ring and its torus Fermi surface.
This generic model allows us to extract analytic results for the frequencies and phase shifts of oscillation (Tab. \ref{tab:freq-pha}) from the calculated resistivity in a magnetic field normal to the nodal-ring plane.
From the analytic results, we summarize the general rules for the phase shift in arbitrary cases (Tab. \ref{tab:rule}). The general rules help us to deal with more complex nodal lines and their corresponding phase shifts, such as those in Cu$_3$PdN (Fig. \ref{fig:CuPdN} and Tab. \ref{tab:CuPdN}) and ZrSiS (Fig. \ref{fig:ZrSiS} and Tab. \ref{tab:ZrSiS}) \cite{singha2017large,ali2016butterfly,wang2016evidence,lv2016extremely,hu2016evidence1,PanH17arXiv}.
These results will help us explore transport signatures for nodal-line semimetals and other topological phases of matter.

{\color{red}\emph{Torus Fermi surface.}}-- A generic Hamiltonian of nodal-line semimetals can be given as \cite{bian2016drumhead,bian2016topological}
\begin{equation}\label{eq:model}
  H=\left\{\left[\hbar^2( k_x^2+ k_y^2)/2 m-u\right]\tau _3+\lambda  k_z \tau _1\right\}\otimes \sigma _0,
\end{equation}
where $\tau$, $\sigma$ are the Pauli matrices, ${\bf k}=(k_x,k_y,k_z)$ is the wave vector, and $m$, $u$ and $\lambda$ are model parameters \cite{bian2016topological,bian2016drumhead}. The model describes four energy bands. The dispersions of the energy bands are
$E_{\pm}=\pm \sqrt{[{\hbar^2(k_x^2+k_y^2)}/{2 m}-u]^2+\lambda^2k_z^2}$. For positive $u$, two bands intersect when $k_x^2+k_y^2=2mu/\hbar^2$ at zero energy describing the nodal-line ring with radius $\sqrt{2mu}/\hbar$ [Fig. \ref{fig:torus}(a)]. When the Fermi energy $E_F<u$, the dispersions give rise to a torus Fermi surface in 3D momentum space [Fig. \ref{fig:torus}(a)]. For $E_F>u$, the Fermi surface becomes a drum-like structure [Fig. \ref{fig:torus}(a)]. We will focus on the torus Fermi surface due to the low carrier density in the experiments \cite{schoop2016dirac,neupane2016observation} and its anisotropy as a signature of nodal-line semimetals.

{\color{red}\emph{Phase shifts in $B_z$.}}--
For conventional electrons and Weyl fermions with spherical Fermi surfaces, the frequency and phase shift in the quantum oscillation can be explored theoretically by calculating the resistivity or using the Fermi surface analysis. However, their equivalence is unknown for the sophisticated Fermi surface of nodal-line semimetals. Below we will focus on a solvable case when the magnetic field is perpendicular to the nodal-line plane. In this way, we will extract the rules of determining the phase shift and apply them to more sophisticated cases.

First, we use the Fermi surface analysis.
According to the Onsager relation $F=(\hbar/2\pi e) A$, the oscillation frequency $F$ is proportional to the extremal cross sectional area $A$ on the Fermi surface that is normal to the magnetic field \cite{Onsager52pm}. In a magnetic field normal to the nodal-line plane, specifically $B_z$ here, the extremal cross sections can be found by using
$\left.\partial A/\partial k_z\right|_{E_F}=0$.
There are two extremal cross sections [Fig. \ref{fig:torus}(b)], which lie at the $k_z=0$ plane, where the energy dispersions reduce to
$E_{\pm}=\pm\left|{\hbar^2(k_x^2+k_y^2)}/{2m}-u\right|$,
for electron ($+$) and hole ($-$), respectively.
Using this dispersion and Onsager relation, the high and low frequencies are found to be  $F_\alpha=m(u+E_F)/\hbar e$ for the outer circle and $F_\beta=m(u-E_F)/\hbar e$ for the inner circle, respectively. The two frequencies are expected to induce a beating pattern in the quantum oscillation, which has been well studied \cite{phillips2014tunable}.

\begin{table}[htbp]
\caption{Phase shifts $\phi$ of quantum oscillations in systems with different dispersions (linear or parabolic) and dimensionalities (2D or 3D). $B_z$ and $B_{||}$ are the magnetic fields out of and in the nodal-line plane. $\alpha$, $\beta$, $\gamma$, and $\delta$ correspond to the cross sections of the torus Fermi surface in Fig. \ref{fig:torus}.}
\begin{ruledtabular}
\begin{tabular}{cccc}
System & Electron carrier & Hole carrier  \\
\hline
2D parabolic & $-1/2$  & 1/2
\\
3D parabolic & $-5/8$   & 5/8
\\
2D linear & 0  & 0
\\
3D linear  & $-1/8$ & 1/8
\\
Nodal-line in $B_z$ & $-5/8$ $(\alpha)$, $5/8$ $(\beta)$   & $5/8$ $(\alpha)$, $-5/8$ $(\beta)$
\\
Nodal-line in $B_{||}$ & $-5/8$ $(\gamma)$, $1/8$ $(\delta)$   & $5/8$ $(\gamma)$, $-1/8$ $(\delta)$
\tabularnewline
\end{tabular}\label{tab:phase}
\end{ruledtabular}
\end{table}

The phase shift for each frequency component of the quantum oscillation can be argued according to the relation
\begin{eqnarray}\label{eq:phase}
\phi=-{1}/{2}+{\phi_{\rm B}}/{2\pi}+\phi_{\rm 3D},
\end{eqnarray}
where $\phi_{\rm B}$ is the Berry phase \cite{Mikitik99prl,Xiao10rmp} and $\phi_{\rm 3D}=\mp1/8$ is a correction that arises only in 3D. In previous systems with spherical Fermi surfaces, the sign of $\phi_{\rm 3D}$ is determined by the curvature of the Fermi surface along the direction of the magnetic field \cite{Lifshitz1956theory,Shoenberg62,Coleridge72jpf,Lukyanchuk04prl}.
Nevertheless, the significance of $\phi_{\rm 3D}$ was not fully appreciated in recent literature.
Here, we summarize simple rules for $\phi_{\rm 3D}$. For a maximum cross section, $\phi_{\rm 3D}=-1/8$ for electron carriers and $1/8$ for hole carriers; For a minimum cross section, $\phi_{\rm 3D}=1/8$ for electron carriers and $-1/8$ for hole carriers. In previous systems where the Fermi surface is a sphere, there is only a maximum, so $\phi_{\rm 3D}=-1/8$ for electron carriers and $1/8$ for hole carriers. A parabolic energy band has no Berry phase, so according to these rules, the phase shift is $\pm 1/2$ and $\pm 5/8$ in 2D and 3D, respectively. In contrast, an energy band with linear dispersion carries an extra $\pi$ Berry phase \cite{Mikitik99prl}, so the phase shift is $0$ \cite{ZhangYB05nat} and $\pm 1/8$ \cite{Murakawa13sci} in 2D and 3D, respectively. We summarize the phase shifts of 2D and 3D bands of linear and parabolic dispersions  in Tab. \ref{tab:phase}.

\begin{table}[htb]
    \centering
\caption{The phase shift $\phi$ of the nodal-line semimetal in Fig. \ref{fig:torus}, obtained by using the relation $\phi=-1/2+\phi_{\rm B}/2\pi+\phi_{\rm 3D}$, where $\phi_{\rm B }$ is the Berry phase and $\phi_{\rm 3D}$ is the dimension correction. $\alpha,\beta,\gamma,\delta$ are the extremal cross sections in Fig. \ref{fig:torus}. Max. or Min. indicates whether the cross section is maximum or minimum. For electron carriers, $\phi_{\rm 3D}$ is $-1/8$ for maximum cross section and $1/8$ for minimum cross section. The phase shifts of hole carriers are opposite to those of electrons.}\label{tab:rule}
\begin{ruledtabular}
\begin{tabular}{ccccc}
     & Berry       & Min.          &  Electron            & Hole      \\
     & phase       &  /Max.   &         &      \\ \hline
$\alpha$          &   0          & Max.  & $-1/2+0-1/8=-5/8$ & $+5/8$\\
$\beta$          &   0          & Min.  & $-1/2+0+1/8=-3/8 \leftrightarrow 5/8$ & $ -5/8$\\
$\gamma$          &   0          & Max.  & $-1/2+0-1/8=-5/8$ & $+5/8$\\
$\delta$          &   $\pi$          & Min.  & $-1/2+\pi/2\pi+1/8=1/8$ & $-1/8$\\
\end{tabular}
\end{ruledtabular}
\end{table}

The phase shift is more complex in the nodal-line semimetal. First, the torus Fermi surface allows both maximum and minimum cross sections, in either in-plane or out-of-plane magnetic fields. Second, the Berry phase is 0 and $\pi$, respectively, along the loop parallel to and around the nodal line. Thus, depending on the magnetic field direction, the quantum oscillation has different phase shifts, as summarized by the last two entries in Tab. \ref{tab:phase}. In a magnetic field ($B_z$) normal to the nodal-line plane, the cross sections are $\alpha$ and $\beta$ in Fig. \ref{fig:torus}. Along circles of $\alpha$ and $\beta$, the Berry phase is 0, so the phase shift should take values as $ 5/8$ or $-5/8$.
Take electron carriers for example (see the first two entries in Tab. \ref{tab:rule}). $\alpha$ is the maximum cross section, so $\phi_{\rm 3D}=-1/8$ and $\phi=-1/2+0-1/8=-5/8$ for electron carriers.
$\beta$ is the minimum cross section, so $\phi_{\rm 3D}=1/8$ and $\phi=-1/2+0+1/8=-3/8$ for electron carriers, which is equivalent to $5/8$ because of the $2\pi$ periodicity of the oscillation.

\begin{table}[htb]
    \centering
\caption{The analytic expressions for the frequencies $F_{\alpha, \beta}$ and phase shifts $\phi_{\alpha, \beta}$ in the resistivity formula Eq. \eqref{rho} for electron carriers. Hole carriers have an extra minus sign in all cases compared to electron carriers. }\label{tab:freq-pha}
\begin{ruledtabular}
\begin{tabular}{ccccc}
                           & \multicolumn{2}{c}{Longitudinal $\rho_{zz}$} & \multicolumn{2}{c}{Transverse $\rho_{xx}$} \\
                     &$\alpha$                 &$\beta$       &$\alpha$     &$\beta$\\ \hline
\rule{0pt}{0.65cm}$F$      & $\frac{m}{\hbar e}(u+E_F)$       & $\frac{m}{\hbar e}(u-E_F)$           &  $\frac{m}{\hbar e}(u+E_F)$            & $\frac{m}{\hbar e}(u-E_F)$      \\
\rule{0pt}{0.65cm}$\phi$             & $-5/8$           & $+5/8 $ & $- 5/8$ & $+5/8$\\
\end{tabular}
\end{ruledtabular}
\end{table}

\emph{\color{red}{Resistivity in $B_z$.}} -- To verify the above argument, we calculate the resistivity (see Supplemental Material \cite{Supp} for details), in the presence of a magnetic field ($B_z$) normal to the nodal-line plane.  $B_z$ quantizes the energy spectrum into a set of 1D bands of Landau levels dispersing with $k_z$,
$ E_{k_z}^{\nu\pm }=\pm \sqrt{\left[\hbar \omega \left(\nu +1/2\right)-u\right]^2+\lambda^2 k_z^2}$,
where $\pm$ denote the upper bands and the lower bands, respectively.
These Landau bands are unique, which can be seen at the band bottoms where $k_z=0$. For the upper bands,
$E_{k_z=0}^{\nu+}=u-\hbar \omega \left(\nu +\frac{1}{2}\right)$ for $0\leq\nu\leq\nu^\ast$ and 
$E_{k_z=0}^{\nu+}=\hbar \omega \left(\nu +\frac{1}{2}\right)-u$ for $\nu\geq\nu^\ast+1$; 
where $\nu^\ast$ is the integer portion of $u/\hbar\omega-{1}/{2}$. For $\nu<\nu^\ast$, the energies of the Landau bands go downward with increasing $\nu$. For $\nu>\nu^\ast$, the bands go upward [see Fig. S2(a) of \cite{Supp}]. This behavior is different from those of usual electrons and Weyl fermions, where the energy is monotonic with the Landau index. According to our resistivity calculations (Ref. \cite{Supp}), the two sets of Landau bands correspond to the two cross sections $\alpha$ and $\beta$, and their upward and downward tendencies correspond to the $\phi_{\rm 3D}=-1/8$ and $+1/8$, respectively.

With the Landau bands in $B_z$, we calculate the resistivities along the $z$ ($\rho_{zz}$, out of the nodel-line plane) and $x$ ($\rho_{xx}$, in the nodel-line plane) directions, respectively, through $\rho_{zz}=1/\sigma_{zz}$ and $\rho_{xx}=\sigma_{yy}/(\sigma_{yy}^2+\sigma_{xy}^2)$, where $\sigma_{zz}$, $\sigma_{yy}$, and $\sigma_{xy}$ are conductivities (Ref. \cite{Supp}).
The calculations yield that for both $\rho_{xx}$ and $\rho_{zz}$, the magnetoresistivities have two terms
\begin{align}\label{rho}
(\rho-\rho_0)/\rho_0&=\mathcal{C_\alpha}\exp(-\lambda_D)\cos[2\pi(F_\alpha/B+\phi_\alpha)]\nonumber\\
&+\mathcal{C_\beta}\exp(-\lambda_D)\cos[2\pi(F_\beta/B+\phi_\beta)],
\end{align}
where the analytic expressions for the oscillation frequencies $F_{\alpha, \beta}$ and the phase shifts $\phi_{\alpha, \beta}$ calculated from the resistivity are listed in Table. \ref{tab:freq-pha}. In this way, we establish the equivalence between the resistivity calculations and the Fermi surface analysis for the phase shifts of quantum oscillation in the nodal-line semimetal.
Similarly, the phase shifts in a magnetic field $B_{||}$ applied in the nodal-line plane [Fig. \ref{fig:torus}(c)] can be found, as listed in Tab. \ref{tab:rule} (details can be found in Sec. S8 of \cite{Supp}). Later we generalize the rules in Tab. \ref{tab:rule} to more realistic and sophisticated cases.

\begin{table}[htb]
\centering
\caption{The phase shifts for the extremal cross sections of Cu$_3$PdN in Fig. \ref{fig:CuPdN}. $m$ and $b$ are the model parameters.}\label{tab:CuPdN}
\begin{ruledtabular}
\begin{tabular}{ccccc}
& Berry  & Min. & Location & Phase shift \\
& phase & /Max. &  & (electron)\\ \hline
$\alpha$ & 0 & Max. & $k_z=0 $ (outer) & $-5/8$\\
$\beta$ & 0 & Min. & $k_z=0 $ (inner) & $-3/8 \leftrightarrow 5/8$\\
$\gamma$ & 0 & Max. & $k_z=\sqrt{(m-E_F)/b}$ & $-5/8$\\
$\delta$ & $\pi$ & Min. & $0<k_z<\sqrt{(m-E_F)/b}$ & $1/8$
\end{tabular}
\end{ruledtabular}
\end{table}

\emph{\color{red}Cu$_3$PdN.} --
A nodal-line semimetal of the Cu$_3$PdN family \cite{yu2015topological} has
three nodal rings orthogonal to each other, at planes $k_x=0$, $k_y=0$, and $k_z=0$, respectively, due to the cubic symmetry.
The effective \textbf{k$\cdot$p} Hamiltonian is
$  H=[m-b(k_x^2+k_y^2+k_z^2)]\tau _3+\lambda  k_xk_yk_z \tau _1$,
where $m$, $b$, and $\lambda$ are model parameters. Fig. \ref{fig:CuPdN}(a) shows the Fermi surface for small $E_F$. In a magnetic field along either the $x$, $y$, or $z$ direction, there are four extremal cross sections as shown in Fig. \ref{fig:CuPdN}(b). Their phase shifts are summarized in Tab. \ref{tab:CuPdN}.

\begin{table}[htb]
\centering
\caption{For the nodal-line semimetal ZrSiS, the phase shifts of the cross sections on the Fermi surface in Fig. \ref{fig:ZrSiS}.}\label{tab:ZrSiS}
\begin{ruledtabular}
\begin{tabular}{cccccc}
& Berry  & Min. & Location & Carrier & Phase\\
& phase & /Max. & & type& shift\\ \hline
$\alpha$ & 0 & Max & $k_z=0 $ (outer) & electron & $-5/8$\\
$\beta$ & 0 & Min & $k_z=0 $ (inner) & electron & $-3/8 \leftrightarrow 5/8$\\
$\gamma$ & $\pi$ & Max & $k_z=0$ & hole & $0$\\
\end{tabular}
\end{ruledtabular}
\end{table}

\emph{\color{red}ZrSiS.} --
Most quantum oscillation experiments were for the ZrSiS family \cite{singha2017large,ali2016butterfly,wang2016evidence,lv2016extremely,hu2016evidence1,PanH17arXiv,
hu2016evidence2,hu2017quantum,kumar2017unusual}, in which electron and hole pockets coexist at the Fermi energy \cite{schoop2016dirac,neupane2016observation,singha2017large,PanH17arXiv}. Fig. \ref{fig:ZrSiS} shows the diamond-shaped electron pockets enclosing the nodal line and the quasi-2D tubular-shaped nodal-line hole pockets at the X points in the Brillouin zone \cite{PanH17arXiv}. Therefore, when the magnetic field is along the $c$ axis (the [001] direction), i.e., perpendicular to the diamond-shaped Fermi surface, there are three extremal cross sections, the outer ($\alpha$) and inner ($\beta$) cross sections of the diamond-shaped electron pocket, and the tubular-shaped hole pockets ($\gamma$). The $\alpha$ and $\beta$ cross sections of the diamond-shaped Fermi surface give phase shifts of $5/8$ and $-5/8$ in the oscillation, respectively. Nevertheless, the frequencies of the $\alpha$ and $\beta$ pockets are too large (about $10^4$ T) to be extracted from the experiments. In contrast, the $\gamma$ nodal-line hole pockets have been identified as the origin of the high frequency (about 210 T) component \cite{PanH17arXiv}. This pocket encloses a nodal line, so it has a $\pi$ Berry phase. Also, it has $\phi_{\rm 3D}=0$ because of its quasi-2D nature. According to Eq. (\ref{eq:phase}), the $\gamma$ pocket has a phase shift of $\phi=-1/2+\pi/2\pi+0=0$, in consistence with the results by Ali \emph{et al.} \cite{ali2016butterfly}.

{\color{red} \emph{Discussion and Perspective}} - To apply the rules of phase shifts, one needs to know the shape of the Fermi surface, carrier type, and whether the extremal cross section encloses a nodal line.
The Fermi surface can be obtained by $k\cdot p$ or tight-binding models. The model in Eq. (\ref{eq:model}) has implied mirror reflection symmetry  \cite{bian2016drumhead,bian2016topological}. Nodal lines are also protected by other symmetries \cite{Fang16cpb,kim2015dirac,Bzduvsek16nature,Alexandradinata1710arXiv}, such as two-fold screw rotation \cite{fang2015topological,Chen2015nc}, four-fold rotation and inversion \cite{Fang16cpb}, non-symmorphic symmetry through a glide plane \cite{schoop2016dirac,neupane2016observation}, etc. Symmetry affects the number and location of nodal lines, but not the properties of an individual nodal line.
The existence of the nodal line guarantees the $\pi$ Berry phase, so the rules of phase shift are irrelevant to which symmetries protect nodal lines, as we have shown the cases of Cu$_3$PdN and ZrSiS as examples.
Even if the nodal line breaks because the protecting symmetry is broken and the conduction and valence bands are separated by finite gap $\Delta$, we can still have a Berry phase for the same integral loop as that of $\Delta=0$, and the Berry phase is corrected to (see details in \cite{Supp})
\begin{align}
\phi_{\rm B}=\pm\pi\left(1-\frac{\Delta}{2E_F}\right),
\end{align}
which shows that quantitative studies are still possible in the presence of a finite gap. In materials, trivial and nodal-line pockets may coexist at the Fermi energy.  If the pockets are well separated, one can first use the fast Fourier transformation analysis to extract distinctive frequency components from the resistivity, then fit the phase shift for each frequency component from Landau fan diagram, and finally compare the data of phase shifts with the theory to identify the nodal lines and trivial bands. In experiments, well-separated pockets can be achieved by tuning the Fermi energy. Recently, new discoveries on quantization rules in oscillations have been found for graphene, 2D materials, topological metals, topological crystalline insulator, Dirac and Weyl semimetals \cite{Alexandradinata17prl,Alexandradinata17arXiv}. Topological contributions have also been found in Bloch oscillations \cite{Holler17arXiv}. Generalizations of these classical notions to nodal-line systems will be topics of fundamental interests in the future. Whether the rules for phase shift can be generalized to extremal orbits shared by electron and hole pockets in type-II Weyl semimetals \cite{Alexandradinata17prl,OBrien16prl} will be an outstanding problem. Quantum oscillations in nodal-line systems have also been addressed recently \cite{YangH18arXiv,Oroszlany18arXiv}.

\emph{Acknowledgements} -- We thank helpful discussions with Hongming Weng. This work was supported by National Basic Research Program of China (Grant No. 2015CB921102), Guangdong Innovative and Entrepreneurial Research Team Program (Grant No. 2016ZT06D348), the National Key R \& D Program (Grant No. 2016YFA0301700), the
National Natural Science Foundation of China (Grants No. 11534001, No. 11525417, No. 11474005, and No. 11574127), and the Science, Technology and Innovation Commission of Shenzhen Municipality (Grant No. ZDSYS20170303165926217). C. L. and C. M. W. contribute equally to this work.


\begin{thebibliography}{75}%
\makeatletter
\providecommand \@ifxundefined [1]{%
 \@ifx{#1\undefined}
}%
\providecommand \@ifnum [1]{%
 \ifnum #1\expandafter \@firstoftwo
 \else \expandafter \@secondoftwo
 \fi
}%
\providecommand \@ifx [1]{%
 \ifx #1\expandafter \@firstoftwo
 \else \expandafter \@secondoftwo
 \fi
}%
\providecommand \natexlab [1]{#1}%
\providecommand \enquote  [1]{``#1''}%
\providecommand \bibnamefont  [1]{#1}%
\providecommand \bibfnamefont [1]{#1}%
\providecommand \citenamefont [1]{#1}%
\providecommand \href@noop [0]{\@secondoftwo}%
\providecommand \href [0]{\begingroup \@sanitize@url \@href}%
\providecommand \@href[1]{\@@startlink{#1}\@@href}%
\providecommand \@@href[1]{\endgroup#1\@@endlink}%
\providecommand \@sanitize@url [0]{\catcode `\\12\catcode `\$12\catcode
  `\&12\catcode `\#12\catcode `\^12\catcode `\_12\catcode `\%12\relax}%
\providecommand \@@startlink[1]{}%
\providecommand \@@endlink[0]{}%
\providecommand \url  [0]{\begingroup\@sanitize@url \@url }%
\providecommand \@url [1]{\endgroup\@href {#1}{\urlprefix }}%
\providecommand \urlprefix  [0]{URL }%
\providecommand \Eprint [0]{\href }%
\providecommand \doibase [0]{http://dx.doi.org/}%
\providecommand \selectlanguage [0]{\@gobble}%
\providecommand \bibinfo  [0]{\@secondoftwo}%
\providecommand \bibfield  [0]{\@secondoftwo}%
\providecommand \translation [1]{[#1]}%
\providecommand \BibitemOpen [0]{}%
\providecommand \bibitemStop [0]{}%
\providecommand \bibitemNoStop [0]{.\EOS\space}%
\providecommand \EOS [0]{\spacefactor3000\relax}%
\providecommand \BibitemShut  [1]{\csname bibitem#1\endcsname}%
\let\auto@bib@innerbib\@empty
\bibitem [{\citenamefont {Burkov}\ \emph {et~al.}(2011)\citenamefont {Burkov},
  \citenamefont {Hook},\ and\ \citenamefont {Balents}}]{Burkov11prb}%
  \BibitemOpen
  \bibfield  {author} {\bibinfo {author} {\bibfnamefont {A.~A.}\ \bibnamefont
  {Burkov}}, \bibinfo {author} {\bibfnamefont {M.~D.}\ \bibnamefont {Hook}}, \
  and\ \bibinfo {author} {\bibfnamefont {L.}~\bibnamefont {Balents}},\
  }\bibfield  {title} {\enquote {\bibinfo {title} {Topological nodal
  semimetals},}\ }\href {\doibase 10.1103/PhysRevB.84.235126} {\bibfield
  {journal} {\bibinfo  {journal} {Phys. Rev. B}\ }\textbf {\bibinfo {volume}
  {84}},\ \bibinfo {pages} {235126} (\bibinfo {year} {2011})}\BibitemShut
  {NoStop}%
\bibitem [{\citenamefont {Chiu}\ and\ \citenamefont
  {Schnyder}(2014)}]{Chiu14prb}%
  \BibitemOpen
  \bibfield  {author} {\bibinfo {author} {\bibfnamefont {C.-K.}\ \bibnamefont
  {Chiu}}\ and\ \bibinfo {author} {\bibfnamefont {A.~P.}\ \bibnamefont
  {Schnyder}},\ }\bibfield  {title} {\enquote {\bibinfo {title} {Classification
  of reflection-symmetry-protected topological semimetals and nodal
  superconductors},}\ }\href {\doibase 10.1103/PhysRevB.90.205136} {\bibfield
  {journal} {\bibinfo  {journal} {Phys. Rev. B}\ }\textbf {\bibinfo {volume}
  {90}},\ \bibinfo {pages} {205136} (\bibinfo {year} {2014})}\BibitemShut
  {NoStop}%
\bibitem [{\citenamefont {Fang}\ \emph {et~al.}(2016)\citenamefont {Fang},
  \citenamefont {Weng}, \citenamefont {Dai},\ and\ \citenamefont
  {Fang}}]{Fang16cpb}%
  \BibitemOpen
  \bibfield  {author} {\bibinfo {author} {\bibfnamefont {C.}~\bibnamefont
  {Fang}}, \bibinfo {author} {\bibfnamefont {H.}~\bibnamefont {Weng}}, \bibinfo
  {author} {\bibfnamefont {X.}~\bibnamefont {Dai}}, \ and\ \bibinfo {author}
  {\bibfnamefont {Z.}~\bibnamefont {Fang}},\ }\bibfield  {title} {\enquote
  {\bibinfo {title} {Topological nodal line semimetals},}\ }\href
  {http://stacks.iop.org/1674-1056/25/i=11/a=117106} {\bibfield  {journal}
  {\bibinfo  {journal} {Chin. Phys. B}\ }\textbf {\bibinfo {volume} {25}},\
  \bibinfo {pages} {117106} (\bibinfo {year} {2016})}\BibitemShut {NoStop}%
\bibitem [{\citenamefont {Yang}\ \emph {et~al.}(2017)\citenamefont {Yang},
  \citenamefont {Bojesen}, \citenamefont {Morimoto},\ and\ \citenamefont
  {Furusaki}}]{Yang17prb}%
  \BibitemOpen
  \bibfield  {author} {\bibinfo {author} {\bibfnamefont {B.-J.}\ \bibnamefont
  {Yang}}, \bibinfo {author} {\bibfnamefont {T.~A.}\ \bibnamefont {Bojesen}},
  \bibinfo {author} {\bibfnamefont {T.}~\bibnamefont {Morimoto}}, \ and\
  \bibinfo {author} {\bibfnamefont {A.}~\bibnamefont {Furusaki}},\ }\bibfield
  {title} {\enquote {\bibinfo {title} {Topological semimetals protected by
  off-centered symmetries in nonsymmorphic crystals},}\ }\href {\doibase
  10.1103/PhysRevB.95.075135} {\bibfield  {journal} {\bibinfo  {journal} {Phys.
  Rev. B}\ }\textbf {\bibinfo {volume} {95}},\ \bibinfo {pages} {075135}
  (\bibinfo {year} {2017})}\BibitemShut {NoStop}%
\bibitem [{\citenamefont {Chen}\ \emph
  {et~al.}(2015{\natexlab{a}})\citenamefont {Chen}, \citenamefont {Xie},
  \citenamefont {Yang}, \citenamefont {Pan}, \citenamefont {Zhang},
  \citenamefont {Cohen},\ and\ \citenamefont {Zhang}}]{Chen15nl}%
  \BibitemOpen
  \bibfield  {author} {\bibinfo {author} {\bibfnamefont {Y.}~\bibnamefont
  {Chen}}, \bibinfo {author} {\bibfnamefont {Y.}~\bibnamefont {Xie}}, \bibinfo
  {author} {\bibfnamefont {S.~A.}\ \bibnamefont {Yang}}, \bibinfo {author}
  {\bibfnamefont {H.}~\bibnamefont {Pan}}, \bibinfo {author} {\bibfnamefont
  {F.}~\bibnamefont {Zhang}}, \bibinfo {author} {\bibfnamefont {M.~L.}\
  \bibnamefont {Cohen}}, \ and\ \bibinfo {author} {\bibfnamefont
  {S.}~\bibnamefont {Zhang}},\ }\bibfield  {title} {\enquote {\bibinfo {title}
  {Nanostructured carbon allotropes with {Weyl}-like loops and points},}\
  }\href {\doibase 10.1021/acs.nanolett.5b02978} {\bibfield  {journal}
  {\bibinfo  {journal} {Nano Letters}\ }\textbf {\bibinfo {volume} {15}},\
  \bibinfo {pages} {6974} (\bibinfo {year} {2015}{\natexlab{a}})}\BibitemShut
  {NoStop}%
\bibitem [{\citenamefont {Bzdu{\v{s}}ek}\ \emph {et~al.}(2016)\citenamefont
  {Bzdu{\v{s}}ek}, \citenamefont {Wu}, \citenamefont {R{\"u}egg}, \citenamefont
  {Sigrist},\ and\ \citenamefont {Soluyanov}}]{Bzduvsek16nature}%
  \BibitemOpen
  \bibfield  {author} {\bibinfo {author} {\bibfnamefont {T.}~\bibnamefont
  {Bzdu{\v{s}}ek}}, \bibinfo {author} {\bibfnamefont {Q.}~\bibnamefont {Wu}},
  \bibinfo {author} {\bibfnamefont {A.}~\bibnamefont {R{\"u}egg}}, \bibinfo
  {author} {\bibfnamefont {M.}~\bibnamefont {Sigrist}}, \ and\ \bibinfo
  {author} {\bibfnamefont {A.~A.}\ \bibnamefont {Soluyanov}},\ }\bibfield
  {title} {\enquote {\bibinfo {title} {Nodal-chain metals},}\ }\href {\doibase
  10.1038/nature19099} {\bibfield  {journal} {\bibinfo  {journal} {Nature}\
  }\textbf {\bibinfo {volume} {538}},\ \bibinfo {pages} {75} (\bibinfo {year}
  {2016})}\BibitemShut {NoStop}%
\bibitem [{\citenamefont {Xu}\ \emph {et~al.}(2011)\citenamefont {Xu},
  \citenamefont {Weng}, \citenamefont {Wang}, \citenamefont {Dai},\ and\
  \citenamefont {Fang}}]{Xu11prl}%
  \BibitemOpen
  \bibfield  {author} {\bibinfo {author} {\bibfnamefont {G.}~\bibnamefont
  {Xu}}, \bibinfo {author} {\bibfnamefont {H.~M.}\ \bibnamefont {Weng}},
  \bibinfo {author} {\bibfnamefont {Z.~J.}\ \bibnamefont {Wang}}, \bibinfo
  {author} {\bibfnamefont {X.}~\bibnamefont {Dai}}, \ and\ \bibinfo {author}
  {\bibfnamefont {Z.}~\bibnamefont {Fang}},\ }\bibfield  {title} {\enquote
  {\bibinfo {title} {Chern semimetal and the quantized anomalous {Hall} effect
  in $\text{HgCr}_2\text{Se}_4$},}\ }\href {\doibase
  10.1103/PhysRevLett.107.186806} {\bibfield  {journal} {\bibinfo  {journal}
  {Phys. Rev. Lett.}\ }\textbf {\bibinfo {volume} {107}},\ \bibinfo {pages}
  {186806} (\bibinfo {year} {2011})}\BibitemShut {NoStop}%
\bibitem [{\citenamefont {Weng}\ \emph {et~al.}(2015)\citenamefont {Weng},
  \citenamefont {Liang}, \citenamefont {Xu}, \citenamefont {Yu}, \citenamefont
  {Fang}, \citenamefont {Dai},\ and\ \citenamefont
  {Kawazoe}}]{weng2015topological}%
  \BibitemOpen
  \bibfield  {author} {\bibinfo {author} {\bibfnamefont {H.}~\bibnamefont
  {Weng}}, \bibinfo {author} {\bibfnamefont {Y.}~\bibnamefont {Liang}},
  \bibinfo {author} {\bibfnamefont {Q.}~\bibnamefont {Xu}}, \bibinfo {author}
  {\bibfnamefont {R.}~\bibnamefont {Yu}}, \bibinfo {author} {\bibfnamefont
  {Z.}~\bibnamefont {Fang}}, \bibinfo {author} {\bibfnamefont {X.}~\bibnamefont
  {Dai}}, \ and\ \bibinfo {author} {\bibfnamefont {Y.}~\bibnamefont
  {Kawazoe}},\ }\bibfield  {title} {\enquote {\bibinfo {title} {Topological
  node-line semimetal in three-dimensional graphene networks},}\ }\href
  {\doibase 10.1103/PhysRevB.92.045108} {\bibfield  {journal} {\bibinfo
  {journal} {Phys. Rev. B}\ }\textbf {\bibinfo {volume} {92}},\ \bibinfo
  {pages} {045108} (\bibinfo {year} {2015})}\BibitemShut {NoStop}%
\bibitem [{\citenamefont {Yu}\ \emph {et~al.}(2015)\citenamefont {Yu},
  \citenamefont {Weng}, \citenamefont {Fang}, \citenamefont {Dai},\ and\
  \citenamefont {Hu}}]{yu2015topological}%
  \BibitemOpen
  \bibfield  {author} {\bibinfo {author} {\bibfnamefont {R.}~\bibnamefont
  {Yu}}, \bibinfo {author} {\bibfnamefont {H.}~\bibnamefont {Weng}}, \bibinfo
  {author} {\bibfnamefont {Z.}~\bibnamefont {Fang}}, \bibinfo {author}
  {\bibfnamefont {X.}~\bibnamefont {Dai}}, \ and\ \bibinfo {author}
  {\bibfnamefont {X.}~\bibnamefont {Hu}},\ }\bibfield  {title} {\enquote
  {\bibinfo {title} {Topological node-line semimetal and dirac semimetal state
  in antiperovskite {Cu$_3$PdN}},}\ }\href {\doibase
  10.1103/PhysRevLett.115.036807} {\bibfield  {journal} {\bibinfo  {journal}
  {Phys. Rev. Lett.}\ }\textbf {\bibinfo {volume} {115}},\ \bibinfo {pages}
  {036807} (\bibinfo {year} {2015})}\BibitemShut {NoStop}%
\bibitem [{\citenamefont {Kim}\ \emph {et~al.}(2015)\citenamefont {Kim},
  \citenamefont {Wieder}, \citenamefont {Kane},\ and\ \citenamefont
  {Rappe}}]{kim2015dirac}%
  \BibitemOpen
  \bibfield  {author} {\bibinfo {author} {\bibfnamefont {Y.}~\bibnamefont
  {Kim}}, \bibinfo {author} {\bibfnamefont {B.~J.}\ \bibnamefont {Wieder}},
  \bibinfo {author} {\bibfnamefont {C.~L.}\ \bibnamefont {Kane}}, \ and\
  \bibinfo {author} {\bibfnamefont {A.~M.}\ \bibnamefont {Rappe}},\ }\bibfield
  {title} {\enquote {\bibinfo {title} {Dirac line nodes in inversion-symmetric
  crystals},}\ }\href {\doibase 10.1103/PhysRevLett.115.036806} {\bibfield
  {journal} {\bibinfo  {journal} {Phys. Rev. Lett.}\ }\textbf {\bibinfo
  {volume} {115}},\ \bibinfo {pages} {036806} (\bibinfo {year}
  {2015})}\BibitemShut {NoStop}%
\bibitem [{\citenamefont {Fang}\ \emph {et~al.}(2015)\citenamefont {Fang},
  \citenamefont {Chen}, \citenamefont {Kee},\ and\ \citenamefont
  {Fu}}]{fang2015topological}%
  \BibitemOpen
  \bibfield  {author} {\bibinfo {author} {\bibfnamefont {C.}~\bibnamefont
  {Fang}}, \bibinfo {author} {\bibfnamefont {Y.}~\bibnamefont {Chen}}, \bibinfo
  {author} {\bibfnamefont {H.-Y.}\ \bibnamefont {Kee}}, \ and\ \bibinfo
  {author} {\bibfnamefont {L.}~\bibnamefont {Fu}},\ }\bibfield  {title}
  {\enquote {\bibinfo {title} {Topological nodal line semimetals with and
  without spin-orbital coupling},}\ }\href {\doibase
  10.1103/PhysRevB.92.081201} {\bibfield  {journal} {\bibinfo  {journal} {Phys.
  Rev. B}\ }\textbf {\bibinfo {volume} {92}},\ \bibinfo {pages} {081201}
  (\bibinfo {year} {2015})}\BibitemShut {NoStop}%
\bibitem [{\citenamefont {Chen}\ \emph
  {et~al.}(2015{\natexlab{b}})\citenamefont {Chen}, \citenamefont {Lu},\ and\
  \citenamefont {Kee}}]{Chen2015nc}%
  \BibitemOpen
  \bibfield  {author} {\bibinfo {author} {\bibfnamefont {Y.}~\bibnamefont
  {Chen}}, \bibinfo {author} {\bibfnamefont {Y.-M.}\ \bibnamefont {Lu}}, \ and\
  \bibinfo {author} {\bibfnamefont {H.-Y.}\ \bibnamefont {Kee}},\ }\bibfield
  {title} {\enquote {\bibinfo {title} {Topological crystalline metal in
  orthorhombic perovskite iridates},}\ }\href {\doibase 10.1038/ncomms7593}
  {\bibfield  {journal} {\bibinfo  {journal} {Nat. Commun.}\ }\textbf {\bibinfo
  {volume} {6}},\ \bibinfo {pages} {6593} (\bibinfo {year}
  {2015}{\natexlab{b}})}\BibitemShut {NoStop}%
\bibitem [{\citenamefont {Bian}\ \emph
  {et~al.}(2016{\natexlab{a}})\citenamefont {Bian}, \citenamefont {Chang},
  \citenamefont {Zheng}, \citenamefont {Velury}, \citenamefont {Xu},
  \citenamefont {Neupert}, \citenamefont {Chiu}, \citenamefont {Huang},
  \citenamefont {Sanchez}, \citenamefont {Belopolski}, \citenamefont
  {Alidoust}, \citenamefont {Chen}, \citenamefont {Chang}, \citenamefont
  {Bansil}, \citenamefont {Jeng}, \citenamefont {Lin},\ and\ \citenamefont
  {Hasan}}]{bian2016drumhead}%
  \BibitemOpen
  \bibfield  {author} {\bibinfo {author} {\bibfnamefont {G.}~\bibnamefont
  {Bian}},  \emph {et~al.},\ }\bibfield  {title} {\enquote {\bibinfo {title}
  {Drumhead surface states and topological nodal-line fermions in
  {TlTaSe$_2$}},}\ }\href {\doibase 10.1103/PhysRevB.93.121113} {\bibfield
  {journal} {\bibinfo  {journal} {Phys. Rev. B}\ }\textbf {\bibinfo {volume}
  {93}},\ \bibinfo {pages} {121113} (\bibinfo {year}
  {2016}{\natexlab{a}})}\BibitemShut {NoStop}%
\bibitem [{\citenamefont {Xie}\ \emph {et~al.}(2015)\citenamefont {Xie},
  \citenamefont {Schoop}, \citenamefont {Seibel}, \citenamefont {Gibson},
  \citenamefont {Xie},\ and\ \citenamefont {Cava}}]{xie2015new}%
  \BibitemOpen
  \bibfield  {author} {\bibinfo {author} {\bibfnamefont {L.~S.}\ \bibnamefont
  {Xie}}, \bibinfo {author} {\bibfnamefont {L.~M.}\ \bibnamefont {Schoop}},
  \bibinfo {author} {\bibfnamefont {E.~M.}\ \bibnamefont {Seibel}}, \bibinfo
  {author} {\bibfnamefont {Q.~D.}\ \bibnamefont {Gibson}}, \bibinfo {author}
  {\bibfnamefont {W.}~\bibnamefont {Xie}}, \ and\ \bibinfo {author}
  {\bibfnamefont {R.~J.}\ \bibnamefont {Cava}},\ }\bibfield  {title} {\enquote
  {\bibinfo {title} {{A new form of Ca$_3$P$_2$ with a ring of Dirac nodes}},}\
  }\href {\doibase 10.1063/1.4926545} {\bibfield  {journal} {\bibinfo
  {journal} {APL Mater.}\ }\textbf {\bibinfo {volume} {3}},\ \bibinfo {pages}
  {083602} (\bibinfo {year} {2015})}\BibitemShut {NoStop}%
\bibitem [{\citenamefont {Chan}\ \emph {et~al.}(2016)\citenamefont {Chan},
  \citenamefont {Chiu}, \citenamefont {Chou},\ and\ \citenamefont
  {Schnyder}}]{chan20163}%
  \BibitemOpen
  \bibfield  {author} {\bibinfo {author} {\bibfnamefont {Y.-H.}\ \bibnamefont
  {Chan}}, \bibinfo {author} {\bibfnamefont {C.-K.}\ \bibnamefont {Chiu}},
  \bibinfo {author} {\bibfnamefont {M.~Y.}\ \bibnamefont {Chou}}, \ and\
  \bibinfo {author} {\bibfnamefont {A.~P.}\ \bibnamefont {Schnyder}},\
  }\bibfield  {title} {\enquote {\bibinfo {title} {{Ca$_3$P$_2$} and other
  topological semimetals with line nodes and drumhead surface states},}\ }\href
  {\doibase 10.1103/PhysRevB.93.205132} {\bibfield  {journal} {\bibinfo
  {journal} {Phys. Rev. B}\ }\textbf {\bibinfo {volume} {93}},\ \bibinfo
  {pages} {205132} (\bibinfo {year} {2016})}\BibitemShut {NoStop}%
\bibitem [{\citenamefont {Du}\ \emph {et~al.}(2017)\citenamefont {Du},
  \citenamefont {Tang}, \citenamefont {Wang}, \citenamefont {Sheng},
  \citenamefont {Kan}, \citenamefont {Duan}, \citenamefont {Savrasov},\ and\
  \citenamefont {Wan}}]{Du17npjqm}%
  \BibitemOpen
  \bibfield  {author} {\bibinfo {author} {\bibfnamefont {Y.}~\bibnamefont
  {Du}}, \bibinfo {author} {\bibfnamefont {F.}~\bibnamefont {Tang}}, \bibinfo
  {author} {\bibfnamefont {D.}~\bibnamefont {Wang}}, \bibinfo {author}
  {\bibfnamefont {L.}~\bibnamefont {Sheng}}, \bibinfo {author} {\bibfnamefont
  {E.-j.}\ \bibnamefont {Kan}}, \bibinfo {author} {\bibfnamefont {C.-G.}\
  \bibnamefont {Duan}}, \bibinfo {author} {\bibfnamefont {S.~Y.}\ \bibnamefont
  {Savrasov}}, \ and\ \bibinfo {author} {\bibfnamefont {X.}~\bibnamefont
  {Wan}},\ }\bibfield  {title} {\enquote {\bibinfo {title} {{CaTe}: a new
  topological node-line and {Dirac} semimetal},}\ }\href
  {https://www.nature.com/articles/s41535-016-0005-4} {\bibfield  {journal}
  {\bibinfo  {journal} {npj Quantum Materials}\ }\textbf {\bibinfo {volume}
  {2}},\ \bibinfo {pages} {3} (\bibinfo {year} {2017})}\BibitemShut {NoStop}%
\bibitem [{\citenamefont {Zhao}\ \emph {et~al.}(2016)\citenamefont {Zhao},
  \citenamefont {Yu}, \citenamefont {Weng},\ and\ \citenamefont
  {Fang}}]{zhao2016topological}%
  \BibitemOpen
  \bibfield  {author} {\bibinfo {author} {\bibfnamefont {J.}~\bibnamefont
  {Zhao}}, \bibinfo {author} {\bibfnamefont {R.}~\bibnamefont {Yu}}, \bibinfo
  {author} {\bibfnamefont {H.}~\bibnamefont {Weng}}, \ and\ \bibinfo {author}
  {\bibfnamefont {Z.}~\bibnamefont {Fang}},\ }\bibfield  {title} {\enquote
  {\bibinfo {title} {Topological node-line semimetal in compressed black
  phosphorus},}\ }\href {\doibase 10.1103/PhysRevB.94.195104} {\bibfield
  {journal} {\bibinfo  {journal} {Phys. Rev. B}\ }\textbf {\bibinfo {volume}
  {94}},\ \bibinfo {pages} {195104} (\bibinfo {year} {2016})}\BibitemShut
  {NoStop}%
\bibitem [{\citenamefont {Yamakage}\ \emph {et~al.}(2015)\citenamefont
  {Yamakage}, \citenamefont {Yamakawa}, \citenamefont {Tanaka},\ and\
  \citenamefont {Okamoto}}]{yamakage2015line}%
  \BibitemOpen
  \bibfield  {author} {\bibinfo {author} {\bibfnamefont {A.}~\bibnamefont
  {Yamakage}}, \bibinfo {author} {\bibfnamefont {Y.}~\bibnamefont {Yamakawa}},
  \bibinfo {author} {\bibfnamefont {Y.}~\bibnamefont {Tanaka}}, \ and\ \bibinfo
  {author} {\bibfnamefont {Y.}~\bibnamefont {Okamoto}},\ }\bibfield  {title}
  {\enquote {\bibinfo {title} {Line-node {Dirac} semimetal and topological
  insulating phase in noncentrosymmetric pnictides {CaAgX (X= P, As)}},}\
  }\href {\doibase 10.7566/JPSJ.85.013708} {\bibfield  {journal} {\bibinfo
  {journal} {J. Phys. Soc. Japan}\ }\textbf {\bibinfo {volume} {85}},\ \bibinfo
  {pages} {013708} (\bibinfo {year} {2015})}\BibitemShut {NoStop}%
\bibitem [{\citenamefont {Xu}\ \emph {et~al.}(2017)\citenamefont {Xu},
  \citenamefont {Yu}, \citenamefont {Fang}, \citenamefont {Dai},\ and\
  \citenamefont {Weng}}]{xu2017topological}%
  \BibitemOpen
  \bibfield  {author} {\bibinfo {author} {\bibfnamefont {Q.}~\bibnamefont
  {Xu}}, \bibinfo {author} {\bibfnamefont {R.}~\bibnamefont {Yu}}, \bibinfo
  {author} {\bibfnamefont {Z.}~\bibnamefont {Fang}}, \bibinfo {author}
  {\bibfnamefont {X.}~\bibnamefont {Dai}}, \ and\ \bibinfo {author}
  {\bibfnamefont {H.}~\bibnamefont {Weng}},\ }\bibfield  {title} {\enquote
  {\bibinfo {title} {{Topological nodal line semimetals in the CaP$_3$ family
  of materials}},}\ }\href {\doibase 10.1103/PhysRevB.95.045136} {\bibfield
  {journal} {\bibinfo  {journal} {Phys. Rev. B}\ }\textbf {\bibinfo {volume}
  {95}},\ \bibinfo {pages} {045136} (\bibinfo {year} {2017})}\BibitemShut
  {NoStop}%
\bibitem [{\citenamefont {Jin}\ \emph {et~al.}(2017)\citenamefont {Jin},
  \citenamefont {Wang}, \citenamefont {Zhao}, \citenamefont {Du}, \citenamefont
  {Zheng}, \citenamefont {Gan}, \citenamefont {Liu}, \citenamefont {Xu},\ and\
  \citenamefont {Tong}}]{jin2017prediction}%
  \BibitemOpen
  \bibfield  {author} {\bibinfo {author} {\bibfnamefont {Y.-J.}\ \bibnamefont
  {Jin}}, \bibinfo {author} {\bibfnamefont {R.}~\bibnamefont {Wang}}, \bibinfo
  {author} {\bibfnamefont {J.-Z.}\ \bibnamefont {Zhao}}, \bibinfo {author}
  {\bibfnamefont {Y.-P.}\ \bibnamefont {Du}}, \bibinfo {author} {\bibfnamefont
  {C.-D.}\ \bibnamefont {Zheng}}, \bibinfo {author} {\bibfnamefont {L.-Y.}\
  \bibnamefont {Gan}}, \bibinfo {author} {\bibfnamefont {J.-F.}\ \bibnamefont
  {Liu}}, \bibinfo {author} {\bibfnamefont {H.}~\bibnamefont {Xu}}, \ and\
  \bibinfo {author} {\bibfnamefont {S.}~\bibnamefont {Tong}},\ }\bibfield
  {title} {\enquote {\bibinfo {title} {The prediction of a family group of
  two-dimensional node-line semimetals},}\ }\href
  {http://pubs.rsc.org/en/content/articlelanding/2017/nr/c7nr03520a#!divAbstract}
  {\bibfield  {journal} {\bibinfo  {journal} {Nanoscale}\ }\textbf {\bibinfo
  {volume} {9}},\ \bibinfo {pages} {13112} (\bibinfo {year}
  {2017})}\BibitemShut {NoStop}%
\bibitem [{\citenamefont {Zhu}\ \emph {et~al.}(2016)\citenamefont {Zhu},
  \citenamefont {Li},\ and\ \citenamefont {Li}}]{Zhu16prb}%
  \BibitemOpen
  \bibfield  {author} {\bibinfo {author} {\bibfnamefont {Z.}~\bibnamefont
  {Zhu}}, \bibinfo {author} {\bibfnamefont {M.}~\bibnamefont {Li}}, \ and\
  \bibinfo {author} {\bibfnamefont {J.}~\bibnamefont {Li}},\ }\bibfield
  {title} {\enquote {\bibinfo {title} {Topological semimetal to insulator
  quantum phase transition in the {Zintl} compounds
  {$\mathrm{B}{\mathrm{a}}_{2}X(X=\mathrm{Si},\mathrm{Ge})$}},}\ }\href
  {\doibase 10.1103/PhysRevB.94.155121} {\bibfield  {journal} {\bibinfo
  {journal} {Phys. Rev. B}\ }\textbf {\bibinfo {volume} {94}},\ \bibinfo
  {pages} {155121} (\bibinfo {year} {2016})}\BibitemShut {NoStop}%
\bibitem [{\citenamefont {Liang}\ \emph {et~al.}(2016)\citenamefont {Liang},
  \citenamefont {Zhou}, \citenamefont {Yu}, \citenamefont {Wang},\ and\
  \citenamefont {Weng}}]{Liang16prb}%
  \BibitemOpen
  \bibfield  {author} {\bibinfo {author} {\bibfnamefont {Q.-F.}\ \bibnamefont
  {Liang}}, \bibinfo {author} {\bibfnamefont {J.}~\bibnamefont {Zhou}},
  \bibinfo {author} {\bibfnamefont {R.}~\bibnamefont {Yu}}, \bibinfo {author}
  {\bibfnamefont {Z.}~\bibnamefont {Wang}}, \ and\ \bibinfo {author}
  {\bibfnamefont {H.}~\bibnamefont {Weng}},\ }\bibfield  {title} {\enquote
  {\bibinfo {title} {Node-surface and node-line fermions from nonsymmorphic
  lattice symmetries},}\ }\href {\doibase 10.1103/PhysRevB.93.085427}
  {\bibfield  {journal} {\bibinfo  {journal} {Phys. Rev. B}\ }\textbf {\bibinfo
  {volume} {93}},\ \bibinfo {pages} {085427} (\bibinfo {year}
  {2016})}\BibitemShut {NoStop}%
\bibitem [{\citenamefont {Zeng}\ \emph {et~al.}(2015)\citenamefont {Zeng},
  \citenamefont {Fang}, \citenamefont {Chang}, \citenamefont {Chen},
  \citenamefont {Hsieh}, \citenamefont {Bansil}, \citenamefont {Lin},\ and\
  \citenamefont {Fu}}]{zeng2015topological}%
  \BibitemOpen
  \bibfield  {author} {\bibinfo {author} {\bibfnamefont {M.}~\bibnamefont
  {Zeng}}, \bibinfo {author} {\bibfnamefont {C.}~\bibnamefont {Fang}}, \bibinfo
  {author} {\bibfnamefont {G.}~\bibnamefont {Chang}}, \bibinfo {author}
  {\bibfnamefont {Y.-A.}\ \bibnamefont {Chen}}, \bibinfo {author}
  {\bibfnamefont {T.}~\bibnamefont {Hsieh}}, \bibinfo {author} {\bibfnamefont
  {A.}~\bibnamefont {Bansil}}, \bibinfo {author} {\bibfnamefont
  {H.}~\bibnamefont {Lin}}, \ and\ \bibinfo {author} {\bibfnamefont
  {L.}~\bibnamefont {Fu}},\ }\bibfield  {title} {\enquote {\bibinfo {title}
  {Topological semimetals and topological insulators in rare earth
  monopnictides},}\ }\href {https://arxiv.org/abs/1504.03492} {\bibfield
  {journal} {\bibinfo  {journal} {arXiv:1504.03492}\ } (\bibinfo {year}
  {2015})}\BibitemShut {NoStop}%
\bibitem [{\citenamefont {Hirayama}\ \emph {et~al.}(2017)\citenamefont
  {Hirayama}, \citenamefont {Okugawa}, \citenamefont {Miyake},\ and\
  \citenamefont {Murakami}}]{Hirayama17nc}%
  \BibitemOpen
  \bibfield  {author} {\bibinfo {author} {\bibfnamefont {M.}~\bibnamefont
  {Hirayama}}, \bibinfo {author} {\bibfnamefont {R.}~\bibnamefont {Okugawa}},
  \bibinfo {author} {\bibfnamefont {T.}~\bibnamefont {Miyake}}, \ and\ \bibinfo
  {author} {\bibfnamefont {S.}~\bibnamefont {Murakami}},\ }\bibfield  {title}
  {\enquote {\bibinfo {title} {Topological {Dirac} nodal lines and surface
  charges in fcc alkaline earth metals},}\ }\href {\doibase
  10.1038/ncomms14022} {\bibfield  {journal} {\bibinfo  {journal} {Nature
  Commun.}\ }\textbf {\bibinfo {volume} {8}},\ \bibinfo {pages} {14022}
  (\bibinfo {year} {2017})}\BibitemShut {NoStop}%
\bibitem [{\citenamefont {Huang}\ \emph {et~al.}(2016)\citenamefont {Huang},
  \citenamefont {Liu}, \citenamefont {Vanderbilt},\ and\ \citenamefont
  {Duan}}]{Huang16prb}%
  \BibitemOpen
  \bibfield  {author} {\bibinfo {author} {\bibfnamefont {H.}~\bibnamefont
  {Huang}}, \bibinfo {author} {\bibfnamefont {J.}~\bibnamefont {Liu}}, \bibinfo
  {author} {\bibfnamefont {D.}~\bibnamefont {Vanderbilt}}, \ and\ \bibinfo
  {author} {\bibfnamefont {W.}~\bibnamefont {Duan}},\ }\bibfield  {title}
  {\enquote {\bibinfo {title} {Topological nodal-line semimetals in
  alkaline-earth stannides, germanides, and silicides},}\ }\href {\doibase
  10.1103/PhysRevB.93.201114} {\bibfield  {journal} {\bibinfo  {journal} {Phys.
  Rev. B}\ }\textbf {\bibinfo {volume} {93}},\ \bibinfo {pages} {201114}
  (\bibinfo {year} {2016})}\BibitemShut {NoStop}%
\bibitem [{\citenamefont {Li}\ \emph {et~al.}(2016)\citenamefont {Li},
  \citenamefont {Ma}, \citenamefont {Cheng}, \citenamefont {Wang},
  \citenamefont {Li}, \citenamefont {Zhang}, \citenamefont {Li},\ and\
  \citenamefont {Chen}}]{Li16prl}%
  \BibitemOpen
  \bibfield  {author} {\bibinfo {author} {\bibfnamefont {R.}~\bibnamefont
  {Li}}, \bibinfo {author} {\bibfnamefont {H.}~\bibnamefont {Ma}}, \bibinfo
  {author} {\bibfnamefont {X.}~\bibnamefont {Cheng}}, \bibinfo {author}
  {\bibfnamefont {S.}~\bibnamefont {Wang}}, \bibinfo {author} {\bibfnamefont
  {D.}~\bibnamefont {Li}}, \bibinfo {author} {\bibfnamefont {Z.}~\bibnamefont
  {Zhang}}, \bibinfo {author} {\bibfnamefont {Y.}~\bibnamefont {Li}}, \ and\
  \bibinfo {author} {\bibfnamefont {X.-Q.}\ \bibnamefont {Chen}},\ }\bibfield
  {title} {\enquote {\bibinfo {title} {Dirac node lines in pure alkali earth
  metals},}\ }\href {\doibase 10.1103/PhysRevLett.117.096401} {\bibfield
  {journal} {\bibinfo  {journal} {Phys. Rev. Lett.}\ }\textbf {\bibinfo
  {volume} {117}},\ \bibinfo {pages} {096401} (\bibinfo {year}
  {2016})}\BibitemShut {NoStop}%
\bibitem [{\citenamefont {Wang}\ \emph
  {et~al.}(2016{\natexlab{a}})\citenamefont {Wang}, \citenamefont {Weng},
  \citenamefont {Nie}, \citenamefont {Fang}, \citenamefont {Kawazoe},\ and\
  \citenamefont {Chen}}]{Wang16prlbody}%
  \BibitemOpen
  \bibfield  {author} {\bibinfo {author} {\bibfnamefont {J.-T.}\ \bibnamefont
  {Wang}}, \bibinfo {author} {\bibfnamefont {H.}~\bibnamefont {Weng}}, \bibinfo
  {author} {\bibfnamefont {S.}~\bibnamefont {Nie}}, \bibinfo {author}
  {\bibfnamefont {Z.}~\bibnamefont {Fang}}, \bibinfo {author} {\bibfnamefont
  {Y.}~\bibnamefont {Kawazoe}}, \ and\ \bibinfo {author} {\bibfnamefont
  {C.}~\bibnamefont {Chen}},\ }\bibfield  {title} {\enquote {\bibinfo {title}
  {Body-centered orthorhombic ${\mathrm{c}}_{16}$: A novel topological
  node-line semimetal},}\ }\href {\doibase 10.1103/PhysRevLett.116.195501}
  {\bibfield  {journal} {\bibinfo  {journal} {Phys. Rev. Lett.}\ }\textbf
  {\bibinfo {volume} {116}},\ \bibinfo {pages} {195501} (\bibinfo {year}
  {2016}{\natexlab{a}})}\BibitemShut {NoStop}%
\bibitem [{\citenamefont {Sun}\ \emph {et~al.}(2017)\citenamefont {Sun},
  \citenamefont {Zhang}, \citenamefont {Liu}, \citenamefont {Felser},\ and\
  \citenamefont {Yan}}]{SunY17prb}%
  \BibitemOpen
  \bibfield  {author} {\bibinfo {author} {\bibfnamefont {Y.}~\bibnamefont
  {Sun}}, \bibinfo {author} {\bibfnamefont {Y.}~\bibnamefont {Zhang}}, \bibinfo
  {author} {\bibfnamefont {C.-X.}\ \bibnamefont {Liu}}, \bibinfo {author}
  {\bibfnamefont {C.}~\bibnamefont {Felser}}, \ and\ \bibinfo {author}
  {\bibfnamefont {B.}~\bibnamefont {Yan}},\ }\bibfield  {title} {\enquote
  {\bibinfo {title} {{Dirac} nodal lines and induced spin {Hall} effect in
  metallic rutile oxides},}\ }\href {\doibase 10.1103/PhysRevB.95.235104}
  {\bibfield  {journal} {\bibinfo  {journal} {Phys. Rev. B}\ }\textbf {\bibinfo
  {volume} {95}},\ \bibinfo {pages} {235104} (\bibinfo {year}
  {2017})}\BibitemShut {NoStop}%
\bibitem [{\citenamefont {Bian}\ \emph
  {et~al.}(2016{\natexlab{b}})\citenamefont {Bian}, \citenamefont {Chang},
  \citenamefont {Sankar}, \citenamefont {Xu}, \citenamefont {Zheng},
  \citenamefont {Neupert}, \citenamefont {Chiu}, \citenamefont {Huang},
  \citenamefont {Chang}, \citenamefont {Belopolski}, \citenamefont {Sanchez},
  \citenamefont {Neupane}, \citenamefont {Alidoust}, \citenamefont {Liu},
  \citenamefont {Wang}, \citenamefont {Lee}, \citenamefont {Jeng},
  \citenamefont {Zhang}, \citenamefont {Yuan}, \citenamefont {Jia},
  \citenamefont {Bansil}, \citenamefont {Chou}, \citenamefont {Lin},\ and\
  \citenamefont {Hasan}}]{bian2016topological}%
  \BibitemOpen
  \bibfield  {author} {\bibinfo {author} {\bibfnamefont {G.}~\bibnamefont
  {Bian}},  \emph {et~al.},\ }\bibfield  {title} {\enquote {\bibinfo {title}
  {Topological nodal-line fermions in spin-orbit metal {PbTaSe$_2$}},}\ }\href
  {\doibase 10.1038/ncomms10556} {\bibfield  {journal} {\bibinfo  {journal}
  {Nat. Commun.}\ }\textbf {\bibinfo {volume} {7}},\ \bibinfo {pages} {10556}
  (\bibinfo {year} {2016}{\natexlab{b}})}\BibitemShut {NoStop}%
\bibitem [{\citenamefont {Schoop}\ \emph {et~al.}(2016)\citenamefont {Schoop},
  \citenamefont {Ali}, \citenamefont {Stra{\ss}er}, \citenamefont {Topp},
  \citenamefont {Varykhalov}, \citenamefont {Marchenko}, \citenamefont
  {Duppel}, \citenamefont {Parkin}, \citenamefont {Lotsch},\ and\ \citenamefont
  {Ast}}]{schoop2016dirac}%
  \BibitemOpen
  \bibfield  {author} {\bibinfo {author} {\bibfnamefont {L.~M.}\ \bibnamefont
  {Schoop}}, \bibinfo {author} {\bibfnamefont {M.~N.}\ \bibnamefont {Ali}},
  \bibinfo {author} {\bibfnamefont {C.}~\bibnamefont {Stra{\ss}er}}, \bibinfo
  {author} {\bibfnamefont {A.}~\bibnamefont {Topp}}, \bibinfo {author}
  {\bibfnamefont {A.}~\bibnamefont {Varykhalov}}, \bibinfo {author}
  {\bibfnamefont {D.}~\bibnamefont {Marchenko}}, \bibinfo {author}
  {\bibfnamefont {V.}~\bibnamefont {Duppel}}, \bibinfo {author} {\bibfnamefont
  {S.~S.~P.}\ \bibnamefont {Parkin}}, \bibinfo {author} {\bibfnamefont {B.~V.}\
  \bibnamefont {Lotsch}}, \ and\ \bibinfo {author} {\bibfnamefont {C.~R.}\
  \bibnamefont {Ast}},\ }\bibfield  {title} {\enquote {\bibinfo {title} {{Dirac
  cone protected by non-symmorphic symmetry and three-dimensional Dirac line
  node in ZrSiS}},}\ }\href {\doibase 10.1038/ncomms11696} {\bibfield
  {journal} {\bibinfo  {journal} {Nat. Commun.}\ }\textbf {\bibinfo {volume}
  {7}},\ \bibinfo {pages} {11696} (\bibinfo {year} {2016})}\BibitemShut
  {NoStop}%
\bibitem [{\citenamefont {Neupane}\ \emph {et~al.}(2016)\citenamefont
  {Neupane}, \citenamefont {Belopolski}, \citenamefont {Hosen}, \citenamefont
  {Sanchez}, \citenamefont {Sankar}, \citenamefont {Szlawska}, \citenamefont
  {Xu}, \citenamefont {Dimitri}, \citenamefont {Dhakal}, \citenamefont
  {Maldonado}, \citenamefont {Oppeneer}, \citenamefont {Kaczorowski},
  \citenamefont {Chou}, \citenamefont {Hasan},\ and\ \citenamefont
  {Durakiewicz}}]{neupane2016observation}%
  \BibitemOpen
  \bibfield  {author} {\bibinfo {author} {\bibfnamefont {M.}~\bibnamefont
  {Neupane}},  \emph {et~al.},\ }\bibfield  {title} {\enquote {\bibinfo {title}
  {{Observation of topological nodal fermion semimetal phase in ZrSiS}},}\
  }\href {\doibase 10.1103/PhysRevB.93.201104} {\bibfield  {journal} {\bibinfo
  {journal} {Phys. Rev. B}\ }\textbf {\bibinfo {volume} {93}},\ \bibinfo
  {pages} {201104} (\bibinfo {year} {2016})}\BibitemShut {NoStop}%
\bibitem [{\citenamefont {Chen}\ \emph {et~al.}(2017)\citenamefont {Chen},
  \citenamefont {Xu}, \citenamefont {Jiang}, \citenamefont {Wu}, \citenamefont
  {Qi}, \citenamefont {Yang}, \citenamefont {Wang}, \citenamefont {Sun},
  \citenamefont {Schr\"oter}, \citenamefont {Yang}, \citenamefont {Schoop},
  \citenamefont {Lv}, \citenamefont {Zhou}, \citenamefont {Chen}, \citenamefont
  {Yao}, \citenamefont {Lu}, \citenamefont {Chen}, \citenamefont {Felser},
  \citenamefont {Yan}, \citenamefont {Liu},\ and\ \citenamefont
  {Chen}}]{ChenC17prb}%
  \BibitemOpen
  \bibfield  {author} {\bibinfo {author} {\bibfnamefont {C.}~\bibnamefont
  {Chen}},  \emph {et~al.},\ }\bibfield  {title} {\enquote {\bibinfo {title}
  {{Dirac} line nodes and effect of spin-orbit coupling in the nonsymmorphic
  critical semimetals {$M$}{SiS}({$M$}={Hf},{Zr})},}\ }\href {\doibase
  10.1103/PhysRevB.95.125126} {\bibfield  {journal} {\bibinfo  {journal} {Phys.
  Rev. B}\ }\textbf {\bibinfo {volume} {95}},\ \bibinfo {pages} {125126}
  (\bibinfo {year} {2017})}\BibitemShut {NoStop}%
\bibitem [{\citenamefont {Wu}\ \emph {et~al.}(2016)\citenamefont {Wu},
  \citenamefont {Wang}, \citenamefont {Mun}, \citenamefont {Johnson},
  \citenamefont {Mou}, \citenamefont {Huang}, \citenamefont {Lee},
  \citenamefont {Bud'ko}, \citenamefont {Canfield},\ and\ \citenamefont
  {Kaminski}}]{wu2016dirac}%
  \BibitemOpen
  \bibfield  {author} {\bibinfo {author} {\bibfnamefont {Y.}~\bibnamefont
  {Wu}}, \bibinfo {author} {\bibfnamefont {L.-L.}\ \bibnamefont {Wang}},
  \bibinfo {author} {\bibfnamefont {E.}~\bibnamefont {Mun}}, \bibinfo {author}
  {\bibfnamefont {D.~D.}\ \bibnamefont {Johnson}}, \bibinfo {author}
  {\bibfnamefont {D.}~\bibnamefont {Mou}}, \bibinfo {author} {\bibfnamefont
  {L.}~\bibnamefont {Huang}}, \bibinfo {author} {\bibfnamefont
  {Y.}~\bibnamefont {Lee}}, \bibinfo {author} {\bibfnamefont {S.~L.}\
  \bibnamefont {Bud'ko}}, \bibinfo {author} {\bibfnamefont {P.~C.}\
  \bibnamefont {Canfield}}, \ and\ \bibinfo {author} {\bibfnamefont
  {A.}~\bibnamefont {Kaminski}},\ }\bibfield  {title} {\enquote {\bibinfo
  {title} {{Dirac node arcs in PtSn$_4$}},}\ }\href {\doibase
  10.1038/nphys3712} {\bibfield  {journal} {\bibinfo  {journal} {Nat. Phys.}\
  }\textbf {\bibinfo {volume} {12}},\ \bibinfo {pages} {667} (\bibinfo {year}
  {2016})}\BibitemShut {NoStop}%
\bibitem [{\citenamefont {Murakawa}\ \emph
  {et~al.}(2013{\natexlab{a}})\citenamefont {Murakawa}, \citenamefont
  {Bahramy}, \citenamefont {Tokunaga}, \citenamefont {Kohama}, \citenamefont
  {Bell}, \citenamefont {Kaneko}, \citenamefont {Nagaosa}, \citenamefont
  {Hwang},\ and\ \citenamefont {Tokura}}]{murakawa2013detection}%
  \BibitemOpen
  \bibfield  {author} {\bibinfo {author} {\bibfnamefont {H.}~\bibnamefont
  {Murakawa}}, \bibinfo {author} {\bibfnamefont {M.~S.}\ \bibnamefont
  {Bahramy}}, \bibinfo {author} {\bibfnamefont {M.}~\bibnamefont {Tokunaga}},
  \bibinfo {author} {\bibfnamefont {Y.}~\bibnamefont {Kohama}}, \bibinfo
  {author} {\bibfnamefont {C.}~\bibnamefont {Bell}}, \bibinfo {author}
  {\bibfnamefont {Y.}~\bibnamefont {Kaneko}}, \bibinfo {author} {\bibfnamefont
  {N.}~\bibnamefont {Nagaosa}}, \bibinfo {author} {\bibfnamefont {H.~Y.}\
  \bibnamefont {Hwang}}, \ and\ \bibinfo {author} {\bibfnamefont
  {Y.}~\bibnamefont {Tokura}},\ }\bibfield  {title} {\enquote {\bibinfo {title}
  {{Detection of Berry's phase in a bulk Rashba semiconductor}},}\ }\href
  {\doibase 10.1126/science.1242247} {\bibfield  {journal} {\bibinfo  {journal}
  {Science}\ }\textbf {\bibinfo {volume} {342}},\ \bibinfo {pages} {1490}
  (\bibinfo {year} {2013}{\natexlab{a}})}\BibitemShut {NoStop}%
\bibitem [{\citenamefont {Wang}\ \emph
  {et~al.}(2016{\natexlab{b}})\citenamefont {Wang}, \citenamefont {Lu},\ and\
  \citenamefont {Shen}}]{WangCM16prl}%
  \BibitemOpen
  \bibfield  {author} {\bibinfo {author} {\bibfnamefont {C.~M.}\ \bibnamefont
  {Wang}}, \bibinfo {author} {\bibfnamefont {H.-Z.}\ \bibnamefont {Lu}}, \ and\
  \bibinfo {author} {\bibfnamefont {S.-Q.}\ \bibnamefont {Shen}},\ }\bibfield
  {title} {\enquote {\bibinfo {title} {Anomalous phase shift of quantum
  oscillations in {3D} topological semimetals},}\ }\href {\doibase
  10.1103/PhysRevLett.117.077201} {\bibfield  {journal} {\bibinfo  {journal}
  {Phys. Rev. Lett.}\ }\textbf {\bibinfo {volume} {117}},\ \bibinfo {pages}
  {077201} (\bibinfo {year} {2016}{\natexlab{b}})}\BibitemShut {NoStop}%
\bibitem [{\citenamefont {He}\ \emph {et~al.}(2014)\citenamefont {He},
  \citenamefont {Hong}, \citenamefont {Dong}, \citenamefont {Pan},
  \citenamefont {Zhang}, \citenamefont {Zhang},\ and\ \citenamefont
  {Li}}]{He14prl}%
  \BibitemOpen
  \bibfield  {author} {\bibinfo {author} {\bibfnamefont {L.~P.}\ \bibnamefont
  {He}}, \bibinfo {author} {\bibfnamefont {X.~C.}\ \bibnamefont {Hong}},
  \bibinfo {author} {\bibfnamefont {J.~K.}\ \bibnamefont {Dong}}, \bibinfo
  {author} {\bibfnamefont {J.}~\bibnamefont {Pan}}, \bibinfo {author}
  {\bibfnamefont {Z.}~\bibnamefont {Zhang}}, \bibinfo {author} {\bibfnamefont
  {J.}~\bibnamefont {Zhang}}, \ and\ \bibinfo {author} {\bibfnamefont {S.~Y.}\
  \bibnamefont {Li}},\ }\bibfield  {title} {\enquote {\bibinfo {title} {Quantum
  transport evidence for the three-dimensional {Dirac} semimetal phase in
  $\text{Cd}_{3}\text{As}_{2}$},}\ }\href {\doibase
  10.1103/PhysRevLett.113.246402} {\bibfield  {journal} {\bibinfo  {journal}
  {Phys. Rev. Lett.}\ }\textbf {\bibinfo {volume} {113}},\ \bibinfo {pages}
  {246402} (\bibinfo {year} {2014})}\BibitemShut {NoStop}%
\bibitem [{\citenamefont {Novak}\ \emph {et~al.}(2015)\citenamefont {Novak},
  \citenamefont {Sasaki}, \citenamefont {Segawa},\ and\ \citenamefont
  {Ando}}]{Novak15prbr}%
  \BibitemOpen
  \bibfield  {author} {\bibinfo {author} {\bibfnamefont {M.}~\bibnamefont
  {Novak}}, \bibinfo {author} {\bibfnamefont {S.}~\bibnamefont {Sasaki}},
  \bibinfo {author} {\bibfnamefont {K.}~\bibnamefont {Segawa}}, \ and\ \bibinfo
  {author} {\bibfnamefont {Y.}~\bibnamefont {Ando}},\ }\bibfield  {title}
  {\enquote {\bibinfo {title} {Large linear magnetoresistance in the {Dirac}
  semimetal $\text{TlBiSSe}$},}\ }\href {\doibase 10.1103/PhysRevB.91.041203}
  {\bibfield  {journal} {\bibinfo  {journal} {Phys. Rev. B}\ }\textbf {\bibinfo
  {volume} {91}},\ \bibinfo {pages} {041203} (\bibinfo {year}
  {2015})}\BibitemShut {NoStop}%
\bibitem [{\citenamefont {Zhao}\ \emph {et~al.}(2015)\citenamefont {Zhao},
  \citenamefont {Liu}, \citenamefont {Zhang}, \citenamefont {Wang},
  \citenamefont {Wang}, \citenamefont {Lin}, \citenamefont {Xing},
  \citenamefont {Lu}, \citenamefont {Liu}, \citenamefont {Wang}, \citenamefont
  {Brombosz}, \citenamefont {Xiao}, \citenamefont {Jia}, \citenamefont {Xie},\
  and\ \citenamefont {Wang}}]{Zhao15prx}%
  \BibitemOpen
  \bibfield  {author} {\bibinfo {author} {\bibfnamefont {Y.~F.}\ \bibnamefont
  {Zhao}},  \emph {et~al.},\ }\bibfield  {title} {\enquote {\bibinfo {title}
  {Anisotropic $\text{Fermi}$ surface and quantum limit transport in high
  mobility three-dimensional {Dirac} semimetal $\text{Cd}_3\text{As}_2$},}\
  }\href {\doibase 10.1103/PhysRevX.5.031037} {\bibfield  {journal} {\bibinfo
  {journal} {Phys. Rev. X}\ }\textbf {\bibinfo {volume} {5}},\ \bibinfo {pages}
  {031037} (\bibinfo {year} {2015})}\BibitemShut {NoStop}%
\bibitem [{\citenamefont {Du}\ \emph {et~al.}(2016)\citenamefont {Du},
  \citenamefont {Wang}, \citenamefont {Chen}, \citenamefont {Mao},
  \citenamefont {Khan}, \citenamefont {Xu}, \citenamefont {Zhou}, \citenamefont
  {Zhang}, \citenamefont {Yang}, \citenamefont {Chen}, \citenamefont {Feng},\
  and\ \citenamefont {Fang}}]{Du16scpma}%
  \BibitemOpen
  \bibfield  {author} {\bibinfo {author} {\bibfnamefont {J.}~\bibnamefont
  {Du}},  \emph {et~al.},\ }\bibfield  {title} {\enquote {\bibinfo {title}
  {{Large unsaturated positive and negative magnetoresistance in Weyl semimetal
  TaP}},}\ }\href {\doibase 10.1007/s11433-016-5798-4} {\bibfield  {journal}
  {\bibinfo  {journal} {Sci. China-Phys. Mech. Astron.}\ }\textbf {\bibinfo
  {volume} {59}},\ \bibinfo {pages} {657406} (\bibinfo {year}
  {2016})}\BibitemShut {NoStop}%
\bibitem [{\citenamefont {Yang}\ \emph {et~al.}(2015)\citenamefont {Yang},
  \citenamefont {Li}, \citenamefont {Wang}, \citenamefont {Zhen},\ and\
  \citenamefont {Xu}}]{YangXJ15arXiv-NbAs}%
  \BibitemOpen
  \bibfield  {author} {\bibinfo {author} {\bibfnamefont {X.}~\bibnamefont
  {Yang}}, \bibinfo {author} {\bibfnamefont {Y.}~\bibnamefont {Li}}, \bibinfo
  {author} {\bibfnamefont {Z.}~\bibnamefont {Wang}}, \bibinfo {author}
  {\bibfnamefont {Y.}~\bibnamefont {Zhen}}, \ and\ \bibinfo {author}
  {\bibfnamefont {Z.-A.}\ \bibnamefont {Xu}},\ }\bibfield  {title} {\enquote
  {\bibinfo {title} {Observation of negative magnetoresistance and nontrivial
  $\pi$ $\text{Berry}$'s phase in $\text{3D}$ {Weyl} semi-metal
  $\text{NbAs}$},}\ }\href {http://arxiv.org/abs/1506.02283v1} {\bibfield
  {journal} {\bibinfo  {journal} {arXiv:1506.02283}\ } (\bibinfo {year}
  {2015})}\BibitemShut {NoStop}%
\bibitem [{\citenamefont {Xiong}\ \emph {et~al.}(2015)\citenamefont {Xiong},
  \citenamefont {Kushwaha}, \citenamefont {Liang}, \citenamefont {Krizan},
  \citenamefont {Hirschberger}, \citenamefont {Wang}, \citenamefont {Cava},\
  and\ \citenamefont {Ong}}]{Xiong15sci}%
  \BibitemOpen
  \bibfield  {author} {\bibinfo {author} {\bibfnamefont {J.}~\bibnamefont
  {Xiong}}, \bibinfo {author} {\bibfnamefont {S.~K.}\ \bibnamefont {Kushwaha}},
  \bibinfo {author} {\bibfnamefont {T.}~\bibnamefont {Liang}}, \bibinfo
  {author} {\bibfnamefont {J.~W.}\ \bibnamefont {Krizan}}, \bibinfo {author}
  {\bibfnamefont {M.}~\bibnamefont {Hirschberger}}, \bibinfo {author}
  {\bibfnamefont {W.}~\bibnamefont {Wang}}, \bibinfo {author} {\bibfnamefont
  {R.~J.}\ \bibnamefont {Cava}}, \ and\ \bibinfo {author} {\bibfnamefont
  {N.~P.}\ \bibnamefont {Ong}},\ }\bibfield  {title} {\enquote {\bibinfo
  {title} {Evidence for the chiral anomaly in the {Dirac} semimetal
  $\text{Na}_3\text{Bi}$},}\ }\href
  {http://www.sciencemag.org/content/early/2015/09/02/science.aac6089.abstract}
  {\bibfield  {journal} {\bibinfo  {journal} {Science}\ }\textbf {\bibinfo
  {volume} {350}},\ \bibinfo {pages} {413} (\bibinfo {year}
  {2015})}\BibitemShut {NoStop}%
\bibitem [{\citenamefont {Cao}\ \emph {et~al.}(2015)\citenamefont {Cao},
  \citenamefont {Liang}, \citenamefont {Zhang}, \citenamefont {Liu},
  \citenamefont {Huang}, \citenamefont {Jin}, \citenamefont {Chen},
  \citenamefont {Wang}, \citenamefont {Wang}, \citenamefont {Zhao},
  \citenamefont {Li}, \citenamefont {Dai}, \citenamefont {Zou}, \citenamefont
  {Xia}, \citenamefont {Li},\ and\ \citenamefont {Xiu}}]{Cao15nc}%
  \BibitemOpen
  \bibfield  {author} {\bibinfo {author} {\bibfnamefont {J.}~\bibnamefont
  {Cao}},  \emph {et~al.},\ }\bibfield  {title} {\enquote {\bibinfo {title}
  {Landau level splitting in $\text{Cd}_3\text{As}_2$ under high magnetic
  fields},}\ }\href
  {http://www.nature.com/ncomms/2015/150713/ncomms8779/full/ncomms8779.html}
  {\bibfield  {journal} {\bibinfo  {journal} {Nature Commun.}\ }\textbf
  {\bibinfo {volume} {6}},\ \bibinfo {pages} {7779} (\bibinfo {year}
  {2015})}\BibitemShut {NoStop}%
\bibitem [{\citenamefont {Zhang}\ \emph {et~al.}(2017)\citenamefont {Zhang},
  \citenamefont {Xu}, \citenamefont {Wang}, \citenamefont {Lin}, \citenamefont
  {Du}, \citenamefont {Guo}, \citenamefont {Lee}, \citenamefont {Lu},
  \citenamefont {Feng}, \citenamefont {Huang}, \citenamefont {Chang},
  \citenamefont {Hsu}, \citenamefont {Liu}, \citenamefont {Lin}, \citenamefont
  {Li}, \citenamefont {Zhang}, \citenamefont {Zhang}, \citenamefont {Xie},
  \citenamefont {Neupert}, \citenamefont {Hasan}, \citenamefont {Lu},
  \citenamefont {Wang},\ and\ \citenamefont {Jia}}]{ZhangCL17np}%
  \BibitemOpen
  \bibfield  {author} {\bibinfo {author} {\bibfnamefont {C.-L.}\ \bibnamefont
  {Zhang}},  \emph {et~al.},\ }\bibfield  {title} {\enquote {\bibinfo {title}
  {Magnetic-tunnelling-induced {Weyl} node annihilation in {TaP}},}\ }\href
  {\doibase 10.1038/nphys4183} {\bibfield  {journal} {\bibinfo  {journal}
  {Nature Phys.}\ }\textbf {\bibinfo {volume} {13}},\ \bibinfo {pages} {979}
  (\bibinfo {year} {2017})}\BibitemShut {NoStop}%
\bibitem [{\citenamefont {Narayanan}\ \emph {et~al.}(2015)\citenamefont
  {Narayanan}, \citenamefont {Watson}, \citenamefont {Blake}, \citenamefont
  {Bruyant}, \citenamefont {Drigo}, \citenamefont {Chen}, \citenamefont
  {Prabhakaran}, \citenamefont {Yan}, \citenamefont {Felser}, \citenamefont
  {Kong}, \citenamefont {Canfield},\ and\ \citenamefont
  {Coldea}}]{Narayanan15prl}%
  \BibitemOpen
  \bibfield  {author} {\bibinfo {author} {\bibfnamefont {A.}~\bibnamefont
  {Narayanan}},  \emph {et~al.},\ }\bibfield  {title} {\enquote {\bibinfo
  {title} {Linear magnetoresistance caused by mobility fluctuations in
  $n$-doped $ \text{Cd}_3\text{As}_2$},}\ }\href {\doibase
  10.1103/PhysRevLett.114.117201} {\bibfield  {journal} {\bibinfo  {journal}
  {Phys. Rev. Lett.}\ }\textbf {\bibinfo {volume} {114}},\ \bibinfo {pages}
  {117201} (\bibinfo {year} {2015})}\BibitemShut {NoStop}%
\bibitem [{\citenamefont {Park}\ \emph {et~al.}(2011)\citenamefont {Park},
  \citenamefont {Lee}, \citenamefont {Wolff-Fabris}, \citenamefont {Koh},
  \citenamefont {Eom}, \citenamefont {Kim}, \citenamefont {Farhan},
  \citenamefont {Jo}, \citenamefont {Kim}, \citenamefont {Shim},\ and\
  \citenamefont {Kim}}]{Park11prl}%
  \BibitemOpen
  \bibfield  {author} {\bibinfo {author} {\bibfnamefont {J.}~\bibnamefont
  {Park}},  \emph {et~al.},\ }\bibfield  {title} {\enquote {\bibinfo {title}
  {Anisotropic {Dirac} $\text{Fermions}$ in a $\text{Bi}$ square net of
  $\text{SrMnBi}_2$},}\ }\href
  {http://journals.aps.org/prl/abstract/10.1103/PhysRevLett.107.126402}
  {\bibfield  {journal} {\bibinfo  {journal} {Phys. Rev. Lett.}\ }\textbf
  {\bibinfo {volume} {107}},\ \bibinfo {pages} {126402} (\bibinfo {year}
  {2011})}\BibitemShut {NoStop}%
\bibitem [{\citenamefont {Xiang}\ \emph {et~al.}(2015)\citenamefont {Xiang},
  \citenamefont {Wang}, \citenamefont {Veldhorst}, \citenamefont {Dou},\ and\
  \citenamefont {Fuhrer}}]{Xiang15prb}%
  \BibitemOpen
  \bibfield  {author} {\bibinfo {author} {\bibfnamefont {F.-X.}\ \bibnamefont
  {Xiang}}, \bibinfo {author} {\bibfnamefont {X.-L.}\ \bibnamefont {Wang}},
  \bibinfo {author} {\bibfnamefont {M.}~\bibnamefont {Veldhorst}}, \bibinfo
  {author} {\bibfnamefont {S.-X.}\ \bibnamefont {Dou}}, \ and\ \bibinfo
  {author} {\bibfnamefont {M.~S.}\ \bibnamefont {Fuhrer}},\ }\bibfield  {title}
  {\enquote {\bibinfo {title} {Observation of topological transition of
  $\text{Fermi}$ surface from a spindle torus to a torus in bulk
  $\text{Rashba}$ spin-split $\text{BiTeCl}$},}\ }\href
  {http://link.aps.org/doi/10.1103/PhysRevB.92.035123} {\bibfield  {journal}
  {\bibinfo  {journal} {Phys. Rev. B}\ }\textbf {\bibinfo {volume} {92}},\
  \bibinfo {pages} {035123} (\bibinfo {year} {2015})}\BibitemShut {NoStop}%
\bibitem [{\citenamefont {Tafti}\ \emph {et~al.}(2016)\citenamefont {Tafti},
  \citenamefont {Gibson}, \citenamefont {Kushwaha}, \citenamefont
  {Haldolaarachchige},\ and\ \citenamefont {Cava}}]{Tafti16np}%
  \BibitemOpen
  \bibfield  {author} {\bibinfo {author} {\bibfnamefont {F.~F.}\ \bibnamefont
  {Tafti}}, \bibinfo {author} {\bibfnamefont {Q.~D.}\ \bibnamefont {Gibson}},
  \bibinfo {author} {\bibfnamefont {S.~K.}\ \bibnamefont {Kushwaha}}, \bibinfo
  {author} {\bibfnamefont {N.}~\bibnamefont {Haldolaarachchige}}, \ and\
  \bibinfo {author} {\bibfnamefont {R.~J.}\ \bibnamefont {Cava}},\ }\bibfield
  {title} {\enquote {\bibinfo {title} {Resistivity plateau and extreme
  magnetoresistance in {LaSb}},}\ }\href
  {http://www.nature.com/nphys/journal/v12/n3/full/nphys3581.html} {\bibfield
  {journal} {\bibinfo  {journal} {Nature Phys.}\ }\textbf {\bibinfo {volume}
  {12}},\ \bibinfo {pages} {272} (\bibinfo {year} {2016})}\BibitemShut
  {NoStop}%
\bibitem [{\citenamefont {Luo}\ \emph {et~al.}(2015)\citenamefont {Luo},
  \citenamefont {Ghimire}, \citenamefont {Wartenbe}, \citenamefont {Choi},
  \citenamefont {Neupane}, \citenamefont {McDonald}, \citenamefont {Bauer},
  \citenamefont {Zhu}, \citenamefont {Thompson},\ and\ \citenamefont
  {Ronning}}]{Luo15prb}%
  \BibitemOpen
  \bibfield  {author} {\bibinfo {author} {\bibfnamefont {Y.}~\bibnamefont
  {Luo}}, \bibinfo {author} {\bibfnamefont {N.~J.}\ \bibnamefont {Ghimire}},
  \bibinfo {author} {\bibfnamefont {M.}~\bibnamefont {Wartenbe}}, \bibinfo
  {author} {\bibfnamefont {H.}~\bibnamefont {Choi}}, \bibinfo {author}
  {\bibfnamefont {M.}~\bibnamefont {Neupane}}, \bibinfo {author} {\bibfnamefont
  {R.~D.}\ \bibnamefont {McDonald}}, \bibinfo {author} {\bibfnamefont {E.~D.}\
  \bibnamefont {Bauer}}, \bibinfo {author} {\bibfnamefont {J.}~\bibnamefont
  {Zhu}}, \bibinfo {author} {\bibfnamefont {J.~D.}\ \bibnamefont {Thompson}}, \
  and\ \bibinfo {author} {\bibfnamefont {F.}~\bibnamefont {Ronning}},\
  }\bibfield  {title} {\enquote {\bibinfo {title} {Electron-hole compensation
  effect between topologically trivial electrons and nontrivial holes in
  nbas},}\ }\href {\doibase 10.1103/PhysRevB.92.205134} {\bibfield  {journal}
  {\bibinfo  {journal} {Phys. Rev. B}\ }\textbf {\bibinfo {volume} {92}},\
  \bibinfo {pages} {205134} (\bibinfo {year} {2015})}\BibitemShut {NoStop}%
\bibitem [{\citenamefont {Singha}\ \emph {et~al.}(2017)\citenamefont {Singha},
  \citenamefont {Pariari}, \citenamefont {Satpati},\ and\ \citenamefont
  {Mandal}}]{singha2017large}%
  \BibitemOpen
  \bibfield  {author} {\bibinfo {author} {\bibfnamefont {R.}~\bibnamefont
  {Singha}}, \bibinfo {author} {\bibfnamefont {A.~K.}\ \bibnamefont {Pariari}},
  \bibinfo {author} {\bibfnamefont {B.}~\bibnamefont {Satpati}}, \ and\
  \bibinfo {author} {\bibfnamefont {P.}~\bibnamefont {Mandal}},\ }\bibfield
  {title} {\enquote {\bibinfo {title} {{Large nonsaturating magnetoresistance
  and signature of nondegenerate Dirac nodes in ZrSiS}},}\ }\href {\doibase
  10.1073/pnas.1618004114} {\bibfield  {journal} {\bibinfo  {journal} {Proc.
  Natl. Acad. Sci. USA}\ }\textbf {\bibinfo {volume} {114}},\ \bibinfo {pages}
  {2468} (\bibinfo {year} {2017})}\BibitemShut {NoStop}%
\bibitem [{\citenamefont {Ali}\ \emph {et~al.}(2016)\citenamefont {Ali},
  \citenamefont {Schoop}, \citenamefont {Garg}, \citenamefont {Lippmann},
  \citenamefont {Lara}, \citenamefont {Lotsch},\ and\ \citenamefont
  {Parkin}}]{ali2016butterfly}%
  \BibitemOpen
  \bibfield  {author} {\bibinfo {author} {\bibfnamefont {M.~N.}\ \bibnamefont
  {Ali}}, \bibinfo {author} {\bibfnamefont {L.~M.}\ \bibnamefont {Schoop}},
  \bibinfo {author} {\bibfnamefont {C.}~\bibnamefont {Garg}}, \bibinfo {author}
  {\bibfnamefont {J.~M.}\ \bibnamefont {Lippmann}}, \bibinfo {author}
  {\bibfnamefont {E.}~\bibnamefont {Lara}}, \bibinfo {author} {\bibfnamefont
  {B.}~\bibnamefont {Lotsch}}, \ and\ \bibinfo {author} {\bibfnamefont
  {S.~S.~P.}\ \bibnamefont {Parkin}},\ }\bibfield  {title} {\enquote {\bibinfo
  {title} {Butterfly magnetoresistance, {quasi-2D Dirac Fermi} surface and
  topological phase transition in {ZrSiS}},}\ }\href {\doibase
  10.1126/sciadv.1601742} {\bibfield  {journal} {\bibinfo  {journal} {Sci.
  Adv.}\ }\textbf {\bibinfo {volume} {2}},\ \bibinfo {pages} {e1601742}
  (\bibinfo {year} {2016})}\BibitemShut {NoStop}%
\bibitem [{\citenamefont {Wang}\ \emph
  {et~al.}(2016{\natexlab{c}})\citenamefont {Wang}, \citenamefont {Pan},
  \citenamefont {Gao}, \citenamefont {Yu}, \citenamefont {Jiang}, \citenamefont
  {Zhang}, \citenamefont {Zuo}, \citenamefont {Zhang}, \citenamefont {Wei},
  \citenamefont {Niu}, \citenamefont {Xia}, \citenamefont {Wan}, \citenamefont
  {Chen}, \citenamefont {Song}, \citenamefont {Xu}, \citenamefont {Wang},
  \citenamefont {Wang},\ and\ \citenamefont {Zhang}}]{wang2016evidence}%
  \BibitemOpen
  \bibfield  {author} {\bibinfo {author} {\bibfnamefont {X.}~\bibnamefont
  {Wang}},  \emph {et~al.},\ }\bibfield  {title} {\enquote {\bibinfo {title}
  {Evidence of both surface and bulk {Dirac} bands and anisotropic
  nonsaturating magnetoresistance in {ZrSiS}},}\ }\href {\doibase
  10.1002/aelm.201600228} {\bibfield  {journal} {\bibinfo  {journal} {Adv.
  Electron. Mater.}\ }\textbf {\bibinfo {volume} {2}},\ \bibinfo {pages}
  {1600228} (\bibinfo {year} {2016}{\natexlab{c}})}\BibitemShut {NoStop}%
\bibitem [{\citenamefont {Lv}\ \emph {et~al.}(2016)\citenamefont {Lv},
  \citenamefont {Zhang}, \citenamefont {Li}, \citenamefont {Yao}, \citenamefont
  {Chen}, \citenamefont {Zhou}, \citenamefont {Zhang}, \citenamefont {Lu},\
  and\ \citenamefont {Chen}}]{lv2016extremely}%
  \BibitemOpen
  \bibfield  {author} {\bibinfo {author} {\bibfnamefont {Y.-Y.}\ \bibnamefont
  {Lv}}, \bibinfo {author} {\bibfnamefont {B.-B.}\ \bibnamefont {Zhang}},
  \bibinfo {author} {\bibfnamefont {X.}~\bibnamefont {Li}}, \bibinfo {author}
  {\bibfnamefont {S.-H.}\ \bibnamefont {Yao}}, \bibinfo {author} {\bibfnamefont
  {Y.~B.}\ \bibnamefont {Chen}}, \bibinfo {author} {\bibfnamefont
  {J.}~\bibnamefont {Zhou}}, \bibinfo {author} {\bibfnamefont {S.-T.}\
  \bibnamefont {Zhang}}, \bibinfo {author} {\bibfnamefont {M.-H.}\ \bibnamefont
  {Lu}}, \ and\ \bibinfo {author} {\bibfnamefont {Y.-F.}\ \bibnamefont
  {Chen}},\ }\bibfield  {title} {\enquote {\bibinfo {title} {Extremely large
  and significantly anisotropic magnetoresistance in {ZrSiS} single
  crystals},}\ }\href {\doibase 10.1063/1.4953772} {\bibfield  {journal}
  {\bibinfo  {journal} {App. Phys. Lett.}\ }\textbf {\bibinfo {volume} {108}},\
  \bibinfo {pages} {244101} (\bibinfo {year} {2016})}\BibitemShut {NoStop}%
\bibitem [{\citenamefont {Hu}\ \emph {et~al.}(2017{\natexlab{a}})\citenamefont
  {Hu}, \citenamefont {Tang}, \citenamefont {Liu}, \citenamefont {Zhu},
  \citenamefont {Wei},\ and\ \citenamefont {Mao}}]{hu2016evidence1}%
  \BibitemOpen
  \bibfield  {author} {\bibinfo {author} {\bibfnamefont {J.}~\bibnamefont
  {Hu}}, \bibinfo {author} {\bibfnamefont {Z.}~\bibnamefont {Tang}}, \bibinfo
  {author} {\bibfnamefont {J.}~\bibnamefont {Liu}}, \bibinfo {author}
  {\bibfnamefont {Y.}~\bibnamefont {Zhu}}, \bibinfo {author} {\bibfnamefont
  {J.}~\bibnamefont {Wei}}, \ and\ \bibinfo {author} {\bibfnamefont
  {Z.}~\bibnamefont {Mao}},\ }\bibfield  {title} {\enquote {\bibinfo {title}
  {{Nearly massless Dirac fermions and strong Zeeman splitting in the
  nodal-line semimetal ZrSiS probed by de Haas--van Alphen quantum
  oscillations}},}\ }\href {\doibase 10.1103/PhysRevB.96.045127} {\bibfield
  {journal} {\bibinfo  {journal} {Phys. Rev. B}\ }\textbf {\bibinfo {volume}
  {96}},\ \bibinfo {pages} {045127} (\bibinfo {year}
  {2017}{\natexlab{a}})}\BibitemShut {NoStop}%
\bibitem [{\citenamefont {Pan}\ \emph {et~al.}(2017)\citenamefont {Pan},
  \citenamefont {Tong}, \citenamefont {Yu}, \citenamefont {Wang}, \citenamefont
  {Fu}, \citenamefont {Zhang}, \citenamefont {Wu}, \citenamefont {Wan},
  \citenamefont {Zhang}, \citenamefont {Wang},\ and\ \citenamefont
  {Song}}]{PanH17arXiv}%
  \BibitemOpen
  \bibfield  {author} {\bibinfo {author} {\bibfnamefont {H.}~\bibnamefont
  {Pan}},  \emph {et~al.},\ }\bibfield  {title} {\enquote {\bibinfo {title}
  {Three-dimensional anisotropic magnetoresistance in the {Dirac} node-line
  material {ZrSiSe}},}\ }\href {https://arxiv.org/abs/1708.02779} {\bibfield
  {journal} {\bibinfo  {journal} {arXiv:1708.02779}\ } (\bibinfo {year}
  {2017})}\BibitemShut {NoStop}%
\bibitem [{\citenamefont {Hu}\ \emph {et~al.}(2016)\citenamefont {Hu},
  \citenamefont {Tang}, \citenamefont {Liu}, \citenamefont {Liu}, \citenamefont
  {Zhu}, \citenamefont {Graf}, \citenamefont {Myhro}, \citenamefont {Tran},
  \citenamefont {Lau}, \citenamefont {Wei},\ and\ \citenamefont
  {Mao}}]{hu2016evidence2}%
  \BibitemOpen
  \bibfield  {author} {\bibinfo {author} {\bibfnamefont {J.}~\bibnamefont
  {Hu}},  \emph {et~al.},\ }\bibfield  {title} {\enquote {\bibinfo {title}
  {Evidence of topological nodal-line fermions in {ZrSiSe and ZrSiTe}},}\
  }\href {\doibase 10.1103/PhysRevLett.117.016602} {\bibfield  {journal}
  {\bibinfo  {journal} {Phys. Rev. Lett.}\ }\textbf {\bibinfo {volume} {117}},\
  \bibinfo {pages} {016602} (\bibinfo {year} {2016})}\BibitemShut {NoStop}%
\bibitem [{\citenamefont {Hu}\ \emph {et~al.}(2017{\natexlab{b}})\citenamefont
  {Hu}, \citenamefont {Zhu}, \citenamefont {Graf}, \citenamefont {Tang},
  \citenamefont {Liu},\ and\ \citenamefont {Mao}}]{hu2017quantum}%
  \BibitemOpen
  \bibfield  {author} {\bibinfo {author} {\bibfnamefont {J.}~\bibnamefont
  {Hu}}, \bibinfo {author} {\bibfnamefont {Y.~L.}\ \bibnamefont {Zhu}},
  \bibinfo {author} {\bibfnamefont {D.}~\bibnamefont {Graf}}, \bibinfo {author}
  {\bibfnamefont {Z.~J.}\ \bibnamefont {Tang}}, \bibinfo {author}
  {\bibfnamefont {J.~Y.}\ \bibnamefont {Liu}}, \ and\ \bibinfo {author}
  {\bibfnamefont {Z.~Q.}\ \bibnamefont {Mao}},\ }\bibfield  {title} {\enquote
  {\bibinfo {title} {{Quantum oscillation studies of the topological semimetal
  candidate ZrGeM (M=S, Se, Te)}},}\ }\href {\doibase
  10.1103/PhysRevB.95.205134} {\bibfield  {journal} {\bibinfo  {journal} {Phys.
  Rev. B}\ }\textbf {\bibinfo {volume} {95}},\ \bibinfo {pages} {205134}
  (\bibinfo {year} {2017}{\natexlab{b}})}\BibitemShut {NoStop}%
\bibitem [{\citenamefont {Kumar}\ \emph {et~al.}(2017)\citenamefont {Kumar},
  \citenamefont {Manna}, \citenamefont {Qi}, \citenamefont {Wu}, \citenamefont
  {Wang}, \citenamefont {Yan}, \citenamefont {Felser},\ and\ \citenamefont
  {Shekhar}}]{kumar2017unusual}%
  \BibitemOpen
  \bibfield  {author} {\bibinfo {author} {\bibfnamefont {N.}~\bibnamefont
  {Kumar}}, \bibinfo {author} {\bibfnamefont {K.}~\bibnamefont {Manna}},
  \bibinfo {author} {\bibfnamefont {Y.}~\bibnamefont {Qi}}, \bibinfo {author}
  {\bibfnamefont {S.-C.}\ \bibnamefont {Wu}}, \bibinfo {author} {\bibfnamefont
  {L.}~\bibnamefont {Wang}}, \bibinfo {author} {\bibfnamefont {B.}~\bibnamefont
  {Yan}}, \bibinfo {author} {\bibfnamefont {C.}~\bibnamefont {Felser}}, \ and\
  \bibinfo {author} {\bibfnamefont {C.}~\bibnamefont {Shekhar}},\ }\bibfield
  {title} {\enquote {\bibinfo {title} {{Unusual magnetotransport from Si-square
  nets in topological semimetal HfSiS}},}\ }\href {\doibase
  10.1103/PhysRevB.95.121109} {\bibfield  {journal} {\bibinfo  {journal} {Phys.
  Rev. B}\ }\textbf {\bibinfo {volume} {95}},\ \bibinfo {pages} {121109}
  (\bibinfo {year} {2017})}\BibitemShut {NoStop}%
\bibitem [{\citenamefont {Onsager}(1952)}]{Onsager52pm}%
  \BibitemOpen
  \bibfield  {author} {\bibinfo {author} {\bibfnamefont {L.}~\bibnamefont
  {Onsager}},\ }\bibfield  {title} {\enquote {\bibinfo {title} {Interpretation
  of the {de Haas-van Alphen} effect},}\ }\href {\doibase
  10.1080/14786440908521019} {\bibfield  {journal} {\bibinfo  {journal}
  {Philos. Mag.}\ }\textbf {\bibinfo {volume} {43}},\ \bibinfo {pages} {1006}
  (\bibinfo {year} {1952})}\BibitemShut {NoStop}%
\bibitem [{\citenamefont {Phillips}\ and\ \citenamefont
  {Aji}(2014)}]{phillips2014tunable}%
  \BibitemOpen
  \bibfield  {author} {\bibinfo {author} {\bibfnamefont {M.}~\bibnamefont
  {Phillips}}\ and\ \bibinfo {author} {\bibfnamefont {V.}~\bibnamefont {Aji}},\
  }\bibfield  {title} {\enquote {\bibinfo {title} {Tunable line node
  semimetals},}\ }\href {\doibase 10.1103/PhysRevB.90.115111} {\bibfield
  {journal} {\bibinfo  {journal} {Phys. Rev. B}\ }\textbf {\bibinfo {volume}
  {90}},\ \bibinfo {pages} {115111} (\bibinfo {year} {2014})}\BibitemShut
  {NoStop}%
\bibitem [{\citenamefont {Mikitik}\ and\ \citenamefont
  {Sharlai}(1999)}]{Mikitik99prl}%
  \BibitemOpen
  \bibfield  {author} {\bibinfo {author} {\bibfnamefont {G.~P.}\ \bibnamefont
  {Mikitik}}\ and\ \bibinfo {author} {\bibfnamefont {Y.~V.}\ \bibnamefont
  {Sharlai}},\ }\bibfield  {title} {\enquote {\bibinfo {title} {Manifestation
  of berry's phase in metal physics},}\ }\href {\doibase
  10.1103/PhysRevLett.82.2147} {\bibfield  {journal} {\bibinfo  {journal}
  {Phys. Rev. Lett.}\ }\textbf {\bibinfo {volume} {82}},\ \bibinfo {pages}
  {2147} (\bibinfo {year} {1999})}\BibitemShut {NoStop}%
\bibitem [{\citenamefont {Xiao}\ \emph {et~al.}(2010)\citenamefont {Xiao},
  \citenamefont {Chang},\ and\ \citenamefont {Niu}}]{Xiao10rmp}%
  \BibitemOpen
  \bibfield  {author} {\bibinfo {author} {\bibfnamefont {D.}~\bibnamefont
  {Xiao}}, \bibinfo {author} {\bibfnamefont {M.~C.}\ \bibnamefont {Chang}}, \
  and\ \bibinfo {author} {\bibfnamefont {Q.}~\bibnamefont {Niu}},\ }\bibfield
  {title} {\enquote {\bibinfo {title} {Berry phase effects on electronic
  properties},}\ }\href {\doibase 10.1103/RevModPhys.82.1959} {\bibfield
  {journal} {\bibinfo  {journal} {Rev. Mod. Phys.}\ }\textbf {\bibinfo {volume}
  {82}},\ \bibinfo {pages} {1959} (\bibinfo {year} {2010})}\BibitemShut
  {NoStop}%
\bibitem [{\citenamefont {Lifshitz}\ and\ \citenamefont
  {Kosevich}(1956)}]{Lifshitz1956theory}%
  \BibitemOpen
  \bibfield  {author} {\bibinfo {author} {\bibfnamefont {I.~M.}\ \bibnamefont
  {Lifshitz}}\ and\ \bibinfo {author} {\bibfnamefont {A.~M.}\ \bibnamefont
  {Kosevich}},\ }\bibfield  {title} {\enquote {\bibinfo {title} {Theory of
  magnetic susceptibility in metals at low temperatures},}\ }\href@noop {}
  {\bibfield  {journal} {\bibinfo  {journal} {Sov. Phys. JETP}\ }\textbf
  {\bibinfo {volume} {2}},\ \bibinfo {pages} {636} (\bibinfo {year}
  {1956})}\BibitemShut {NoStop}%
\bibitem [{\citenamefont {Shoenberg}(1962)}]{Shoenberg62}%
  \BibitemOpen
  \bibfield  {author} {\bibinfo {author} {\bibfnamefont {D.}~\bibnamefont
  {Shoenberg}},\ }\bibfield  {title} {\enquote {\bibinfo {title} {The {Fermi}
  surfaces of copper, silver and gold. i. the de {Haas}-van {Alphen} effect},}\
  }\href {http://rsta.royalsocietypublishing.org/content/255/1052/85}
  {\bibfield  {journal} {\bibinfo  {journal} {Phil. Trans. R. Soc. Lond. A}\
  }\textbf {\bibinfo {volume} {255}},\ \bibinfo {pages} {85} (\bibinfo {year}
  {1962})}\BibitemShut {NoStop}%
\bibitem [{\citenamefont {Coleridge}\ and\ \citenamefont
  {Templeton}(1972)}]{Coleridge72jpf}%
  \BibitemOpen
  \bibfield  {author} {\bibinfo {author} {\bibfnamefont {P.~T.}\ \bibnamefont
  {Coleridge}}\ and\ \bibinfo {author} {\bibfnamefont {I.~M.}\ \bibnamefont
  {Templeton}},\ }\bibfield  {title} {\enquote {\bibinfo {title} {High
  precision de {Haas-van Alphen} measurements in the noble metals},}\ }\href
  {http://stacks.iop.org/0305-4608/2/i=4/a=009} {\bibfield  {journal} {\bibinfo
   {journal} {J. Phys. F: Metal Phys.}\ }\textbf {\bibinfo {volume} {2}},\
  \bibinfo {pages} {643} (\bibinfo {year} {1972})}\BibitemShut {NoStop}%
\bibitem [{\citenamefont {Luk'yanchuk}\ and\ \citenamefont
  {Kopelevich}(2004)}]{Lukyanchuk04prl}%
  \BibitemOpen
  \bibfield  {author} {\bibinfo {author} {\bibfnamefont {I.~A.}\ \bibnamefont
  {Luk'yanchuk}}\ and\ \bibinfo {author} {\bibfnamefont {Y.}~\bibnamefont
  {Kopelevich}},\ }\bibfield  {title} {\enquote {\bibinfo {title} {Phase
  analysis of quantum oscillations in graphite},}\ }\href {\doibase
  10.1103/PhysRevLett.93.166402} {\bibfield  {journal} {\bibinfo  {journal}
  {Phys. Rev. Lett.}\ }\textbf {\bibinfo {volume} {93}},\ \bibinfo {pages}
  {166402} (\bibinfo {year} {2004})}\BibitemShut {NoStop}%
\bibitem [{\citenamefont {Zhang}\ \emph {et~al.}(2005)\citenamefont {Zhang},
  \citenamefont {Tan}, \citenamefont {Stormer},\ and\ \citenamefont
  {Kim}}]{ZhangYB05nat}%
  \BibitemOpen
  \bibfield  {author} {\bibinfo {author} {\bibfnamefont {Y.}~\bibnamefont
  {Zhang}}, \bibinfo {author} {\bibfnamefont {Y.-W.}\ \bibnamefont {Tan}},
  \bibinfo {author} {\bibfnamefont {H.~L.}\ \bibnamefont {Stormer}}, \ and\
  \bibinfo {author} {\bibfnamefont {P.}~\bibnamefont {Kim}},\ }\bibfield
  {title} {\enquote {\bibinfo {title} {Experimental observation of the quantum
  {Hall} effect and {Berry's} phase in graphene},}\ }\href {\doibase
  10.1038/nature04235} {\bibfield  {journal} {\bibinfo  {journal} {Nature}\
  }\textbf {\bibinfo {volume} {438}},\ \bibinfo {pages} {201} (\bibinfo {year}
  {2005})}\BibitemShut {NoStop}%
\bibitem [{\citenamefont {Murakawa}\ \emph
  {et~al.}(2013{\natexlab{b}})\citenamefont {Murakawa}, \citenamefont
  {Bahramy}, \citenamefont {Tokunaga}, \citenamefont {Kohama}, \citenamefont
  {Bell}, \citenamefont {Kaneko}, \citenamefont {Nagaosa}, \citenamefont
  {Hwang},\ and\ \citenamefont {Tokura}}]{Murakawa13sci}%
  \BibitemOpen
  \bibfield  {author} {\bibinfo {author} {\bibfnamefont {H.}~\bibnamefont
  {Murakawa}}, \bibinfo {author} {\bibfnamefont {M.~S.}\ \bibnamefont
  {Bahramy}}, \bibinfo {author} {\bibfnamefont {M.}~\bibnamefont {Tokunaga}},
  \bibinfo {author} {\bibfnamefont {Y.}~\bibnamefont {Kohama}}, \bibinfo
  {author} {\bibfnamefont {C.}~\bibnamefont {Bell}}, \bibinfo {author}
  {\bibfnamefont {Y.}~\bibnamefont {Kaneko}}, \bibinfo {author} {\bibfnamefont
  {N.}~\bibnamefont {Nagaosa}}, \bibinfo {author} {\bibfnamefont {H.~Y.}\
  \bibnamefont {Hwang}}, \ and\ \bibinfo {author} {\bibfnamefont
  {Y.}~\bibnamefont {Tokura}},\ }\bibfield  {title} {\enquote {\bibinfo {title}
  {Detection of Berry's phase in a bulk Rashba semiconductor},}\ }\href {\doibase 10.1126/science.1242247} {\bibfield
  {journal} {\bibinfo  {journal} {Science}\ }\textbf {\bibinfo {volume}
  {342}},\ \bibinfo {pages} {1490} (\bibinfo {year}
  {2013}{\natexlab{b}})}\BibitemShut {NoStop}%
\bibitem [{Sup()}]{Supp}%
  \BibitemOpen
  \href@noop {} {\bibinfo  {journal} {See Supplemental Material at [url] for detailed calculations, which includes Refs. \cite{Charbonneau82jmp, Vasilopoulos84jmp, Lu2015prb, Zhang2016njp, Vaskospringer2006}}\ }\BibitemShut
  {NoStop}%
\bibitem [{\citenamefont {Alexandradinata}\ and\ \citenamefont
  {Glazman}(2017{\natexlab{a}})}]{Alexandradinata1710arXiv}%
  \BibitemOpen
\bibfield  {journal} {  }\bibfield  {author} {\bibinfo {author} {\bibfnamefont
  {A.}~\bibnamefont {Alexandradinata}}\ and\ \bibinfo {author} {\bibfnamefont
  {L.}~\bibnamefont {Glazman}},\ }\bibfield  {title} {\enquote {\bibinfo
  {title} {Modern semiclassical theory of magnetic transport and breakdown},}\
  }\href {http://arxiv.org/abs/1710.04215} {\bibfield  {journal} {\bibinfo
  {journal} {arXiv:1710.04215}\ } (\bibinfo {year}
  {2017}{\natexlab{a}})}\BibitemShut {NoStop}%
\bibitem [{\citenamefont {Alexandradinata}\ and\ \citenamefont
  {Glazman}(2017{\natexlab{b}})}]{Alexandradinata17prl}%
  \BibitemOpen
  \bibfield  {author} {\bibinfo {author} {\bibfnamefont {A.}~\bibnamefont
  {Alexandradinata}}\ and\ \bibinfo {author} {\bibfnamefont {L.}~\bibnamefont
  {Glazman}},\ }\bibfield  {title} {\enquote {\bibinfo {title} {Geometric phase
  and orbital moment in quantization rules for magnetic breakdown},}\ }\href
  {\doibase 10.1103/PhysRevLett.119.256601} {\bibfield  {journal} {\bibinfo
  {journal} {Phys. Rev. Lett.}\ }\textbf {\bibinfo {volume} {119}},\ \bibinfo
  {pages} {256601} (\bibinfo {year} {2017}{\natexlab{b}})}\BibitemShut
  {NoStop}%
\bibitem [{\citenamefont {Alexandradinata}\ \emph {et~al.}(2017)\citenamefont
  {Alexandradinata}, \citenamefont {Wang}, \citenamefont {Duan},\ and\
  \citenamefont {Glazman}}]{Alexandradinata17arXiv}%
  \BibitemOpen
  \bibfield  {author} {\bibinfo {author} {\bibfnamefont {A.}~\bibnamefont
  {Alexandradinata}}, \bibinfo {author} {\bibfnamefont {C.}~\bibnamefont
  {Wang}}, \bibinfo {author} {\bibfnamefont {W.}~\bibnamefont {Duan}}, \ and\
  \bibinfo {author} {\bibfnamefont {L.}~\bibnamefont {Glazman}},\ }\bibfield
  {title} {\enquote {\bibinfo {title} {Topo-fermiology},}\ }\href
  {http://arxiv.org/abs/1707.08586} {\bibfield  {journal} {\bibinfo  {journal}
  {arXiv:1707.08586}\ } (\bibinfo {year} {2017})}\BibitemShut {NoStop}%
\bibitem [{\citenamefont {H\"{o}ller}\ and\ \citenamefont
  {Alexandradinata}(2017)}]{Holler17arXiv}%
  \BibitemOpen
  \bibfield  {author} {\bibinfo {author} {\bibfnamefont {J.}~\bibnamefont
  {H\"{o}ller}}\ and\ \bibinfo {author} {\bibfnamefont {A.}~\bibnamefont
  {Alexandradinata}},\ }\bibfield  {title} {\enquote {\bibinfo {title}
  {Topological {Bloch} oscillations},}\ }\href
  {http://arxiv.org/abs/1708.02943} {\bibfield  {journal} {\bibinfo  {journal}
  {arXiv:1708.02943}\ } (\bibinfo {year} {2017})}\BibitemShut {NoStop}%
\bibitem [{\citenamefont {O'Brien}\ \emph {et~al.}(2016)\citenamefont
  {O'Brien}, \citenamefont {Diez},\ and\ \citenamefont
  {Beenakker}}]{OBrien16prl}%
  \BibitemOpen
  \bibfield  {author} {\bibinfo {author} {\bibfnamefont {T.~E.}\ \bibnamefont
  {O'Brien}}, \bibinfo {author} {\bibfnamefont {M.}~\bibnamefont {Diez}}, \
  and\ \bibinfo {author} {\bibfnamefont {C.~W.~J.}\ \bibnamefont {Beenakker}},\
  }\bibfield  {title} {\enquote {\bibinfo {title} {Magnetic breakdown and
  {Klein} tunneling in a type-{II} {Weyl} semimetal},}\ }\href {\doibase
  10.1103/PhysRevLett.116.236401} {\bibfield  {journal} {\bibinfo  {journal}
  {Phys. Rev. Lett.}\ }\textbf {\bibinfo {volume} {116}},\ \bibinfo {pages}
  {236401} (\bibinfo {year} {2016})}\BibitemShut {NoStop}%
\bibitem [{\citenamefont {Yang}\ \emph {et~al.}(2018)\citenamefont {Yang},
  \citenamefont {Moessner},\ and\ \citenamefont {Lim}}]{YangH18arXiv}%
  \BibitemOpen
  \bibfield  {author} {\bibinfo {author} {\bibfnamefont {H.}~\bibnamefont
  {Yang}}, \bibinfo {author} {\bibfnamefont {R.}~\bibnamefont {Moessner}}, \
  and\ \bibinfo {author} {\bibfnamefont {L.-K.}\ \bibnamefont {Lim}},\
  }\bibfield  {title} {\enquote {\bibinfo {title} {Quantum oscillations in
  nodal line systems},}\ }\href {http://arxiv.org/abs/1801.02733} {\bibfield
  {journal} {\bibinfo  {journal} {arXiv:1801.02733}\ } (\bibinfo {year}
  {2018})}\BibitemShut {NoStop}%
\bibitem [{\citenamefont {Oroszlany}\ \emph {et~al.}(2018)\citenamefont
  {Oroszlany}, \citenamefont {Dora}, \citenamefont {Cserti},\ and\
  \citenamefont {Cortijo}}]{Oroszlany18arXiv}%
  \BibitemOpen
  \bibfield  {author} {\bibinfo {author} {\bibfnamefont {L.}~\bibnamefont
  {Oroszlany}}, \bibinfo {author} {\bibfnamefont {B.}~\bibnamefont {Dora}},
  \bibinfo {author} {\bibfnamefont {J.}~\bibnamefont {Cserti}}, \ and\ \bibinfo
  {author} {\bibfnamefont {A.}~\bibnamefont {Cortijo}},\ }\bibfield  {title}
  {\enquote {\bibinfo {title} {Topological and trivial magnetic oscillations in
  nodal loop semimetals},}\ }\href {http://arxiv.org/abs/1801.04721} {\bibfield
   {journal} {\bibinfo  {journal} {arXiv:1801.04721}\ } (\bibinfo {year}
  {2018})}\BibitemShut {NoStop}%
\bibitem [{\citenamefont {Charbonneau}\ \emph {et~al.}(1982)\citenamefont
  {Charbonneau}, \citenamefont {Van Vliet},\ and\
  \citenamefont {Vasilopoulos}}]{Charbonneau82jmp}%
  \BibitemOpen
  \bibfield  {author} {\bibinfo {author} {\bibfnamefont {M.}~\bibnamefont
  {Charbonneau}}, \bibinfo {author} {\bibfnamefont {K. M.}~\bibnamefont {Van Vliet}},\ and\ \bibinfo
  {author} {\bibfnamefont {P.}~\bibnamefont {Vasilopoulos}},\ }\bibfield  {title}
  {\enquote {\bibinfo {title} {Linear response theory revisited III: One-body response formulas and generalized Boltzmann equations},}\ }\href {http://aip.scitation.org/doi/10.1063/1.525355} {\bibfield  {journal} {\bibinfo  {journal}
  {J. Math. Phys.}\ }\textbf {\bibinfo {volume} {23}},\ \bibinfo {pages}
  {318} (\bibinfo {year} {1982})}\BibitemShut {NoStop}%
\bibitem [{\citenamefont {Vasilopoulos}\ \emph {et~al.}(1984)\citenamefont
  {Vasilopoulos},\ and\
  \citenamefont {Van Vliet}}]{Vasilopoulos84jmp}%
  \BibitemOpen
  \bibfield  {author} {\bibinfo {author} {\bibfnamefont {P.}~\bibnamefont
  {Vasilopoulos}},\ and\ \bibinfo
  {author} {\bibfnamefont {C. M.}~\bibnamefont {Van Vliet}},\ }\bibfield  {title}
  {\enquote {\bibinfo {title} {Linear response
theory revisited. IV. Applications},}\ }\href {http://aip.scitation.org/doi/10.1063/1.526309} {\bibfield  {journal} {\bibinfo  {journal}
  {J. Math. Phys.}\ }\textbf {\bibinfo {volume} {25}},\ \bibinfo {pages}
  {1391} (\bibinfo {year} {1984})}\BibitemShut {NoStop}%
\bibitem [{\citenamefont {Lu}, \citenamefont {Zhang},\ and\ \citenamefont
  {Shen}(2015)}]{Lu2015prb}%
  \BibitemOpen
  \bibfield  {author} {\bibinfo {author} {\bibfnamefont {H. Z.}~\bibnamefont
  {Lu}}, \bibinfo {author} {\bibfnamefont {S. B.}~\bibnamefont {Zhang}},\ and\ \bibinfo
  {author} {\bibfnamefont {S.-Q.}~\bibnamefont {Shen}},\
  }\bibfield  {title} {\enquote {\bibinfo {title} {High-field magnetoconductivity of topological semimetals with short-range potential},}\ }\href {\doibase 10.1103/PhysRevB.92.045203} {\bibfield
  {journal} {\bibinfo  {journal} {Phys. Rev. B}\ }\textbf {\bibinfo {volume}
  {92}},\ \bibinfo {pages} {045203} (\bibinfo {year} {2015})}\BibitemShut
  {NoStop}%
\bibitem [{\citenamefont {Zhang}, \citenamefont {Lu},\ and\ \citenamefont
  {Shen}(2016)}]{Zhang2016njp}%
  \BibitemOpen
  \bibfield  {author} {\bibinfo {author} {\bibfnamefont {S. B.}~\bibnamefont {Zhang}}, \bibinfo {author} {\bibfnamefont {H. Z.}~\bibnamefont
  {Lu}},\ and\ \bibinfo
  {author} {\bibfnamefont {S.-Q.}~\bibnamefont {Shen}},\
  }\bibfield  {title} {\enquote {\bibinfo {title} {Linear magneto-
conductivity in an intrinsic topological Weyl semimetal},}\ }\href {http://iopscience.iop.org/article/10.1088/1367-2630/18/5/053039/meta} {\bibfield
  {journal} {\bibinfo  {journal} {New J. Phys.}\ }\textbf {\bibinfo {volume}
  {18}},\ \bibinfo {pages} {053039} (\bibinfo {year} {2016})}\BibitemShut
  {NoStop}%
\bibitem{Vaskospringer2006} 
F. T. Vasko and O. E. Raichev,
\textit{Quantum Kinetic Theory and Applications: Electrons, Photons, Phonons} (Springer Science \& Business Media, 2006)\BibitemShut
  {NoStop}%
\end{thebibliography}

%

\end{document}